\documentclass[12pt,reqno]{article}
\usepackage{amsmath}
\usepackage{amsfonts}
\usepackage{amssymb}
\usepackage{latexsym}
\usepackage[dvips]{graphicx}

\usepackage{subfig}
\usepackage{epsf}
\usepackage{multirow}

\allowdisplaybreaks \numberwithin{equation}{section}

\newcommand{\be}{\begin{equation}}
\newcommand{\ee}{\end{equation}}
\newcommand{\bea}{\begin{eqnarray}}
\newcommand{\eea}{\end{eqnarray}}

\let\a=\alpha \let\b=\beta  \let\g=\gamma  
        \let\l=\lambda
\let\m=\mu    \let\n=\nu

\let\G=\Gamma     
         
  \let\eps=\epsilon

\let\==\equiv

\newcommand{\tr}{{\rm tr}}
\newcommand{\Tr}{{\rm Tr}}

\newcommand{\cD}{\mathcal{D}}

\newcommand{\cM}{\mathcal{M}}

\newcommand{\cO}{\mathcal{O}}
\newcommand{\cR}{\mathcal{R}}
\newcommand{\cS}{\mathcal{S}}

\newcommand{\Pt}{\tilde{\Phi}}

\newcommand{\unit}{{\bf{1}}}

\newcommand{\half}{\tfrac{1}{2}}

\newcommand{\gb}{\bar{g}}
\newcommand{\hb}{\bar{h}}

\newcommand{\p}{\partial}

\newcommand{\Cb}{\bar{C}}
\newcommand{\Db}{\bar{D}}

\newcommand{\Pb}{\bar{P}}
\newcommand{\Rb}{\bar{R}}

\newcommand{\rmd}{{\rm d}}

\newcommand{\Dz}{\Delta}

\begin{document}
\thispagestyle{empty}
\begin{flushright} \small
MZ-TH/10-19
\end{flushright}
\bigskip

\begin{center}
 {\LARGE\bfseries  
Bimetric Renormalization Group Flows in \\[1.5ex]
Quantum Einstein Gravity
}
\\[10mm]
Elisa Manrique, Martin Reuter and Frank Saueressig \\[3mm]
{\small\slshape
Institute of Physics, University of Mainz\\
Staudingerweg 7, D-55099 Mainz, Germany \\[1.1ex]
{\upshape\ttfamily manrique@thep.physik.uni-mainz.de} \\
{\upshape\ttfamily reuter@thep.physik.uni-mainz.de} \\
{\upshape\ttfamily saueressig@thep.physik.uni-mainz.de} }\\
\end{center}
\vspace{5mm}

\hrule\bigskip

\centerline{\bfseries Abstract} \medskip \noindent
The formulation of an exact functional renormalization group equation for Quantum Einstein Gravity necessitates 
that the underlying effective average action depends on two metrics, a dynamical metric giving the 
vacuum expectation value of the quantum field, and a background metric supplying the coarse graining scale.
The central requirement of ``background independence'' is met by leaving the background metric completely 
arbitrary. 
This bimetric structure entails that the effective average action may contain three classes 
of interactions: those built from the dynamical metric only, terms which are purely background, 
and those involving a mixture of both metrics.
 This work initiates the first study of the full-fledged gravitational RG flow, which explicitly 
accounts for this bimetric structure, by considering an ansatz for the effective average action which includes 
 all three classes of interactions. It is shown that the non-trivial gravitational RG fixed point
 central to the Asymptotic Safety program  
persists upon disentangling the dynamical and background terms.
 Moreover, upon including the mixed terms, a second non-trivial fixed point emerges, which may 
control the theory's IR behavior. 
\noindent

\bigskip
\hrule\bigskip

\newpage
%
\section{Introduction}
Background independence constitutes one of the central guiding principles
in the quest for a viable quantum theory of gravity. 
This requirement is central in loop quantum gravity \cite{A,R,T} and also implemented in 
lattice approaches towards quantum gravity \cite{hamber}-\cite{ajl3}.
Loosely speaking,
it implies that the spacetime structure realized in Nature should not be part
of the theory's definition, but rather emerge from a dynamical principle. This
strict background invariance, referring to no background structure whatsoever, is, however,
very hard to implement. In particular, without an {\it ab initio} metric the notions of
causality and equal time commutation relations are not defined, so that the usual quantization procedures
underlying ordinary quantum field theories cannot be applied straightforwardly.

A milder, but nevertheless equally admissible road towards a viable quantum gravity theory is the requirement of ``background covariance''.
This allows to introduce a background metric as an auxiliary tool, as long as  none of the theory's basic rules
and assumptions, calculational methods, and predictions, depend on this special metric. In other words, all metrics of physical relevance 
are obtained from the dynamics of the theory. This is the viewpoint adopted in many continuum field theory approaches to quantum gravity, in particular by the functional renormalization group approach initiated in \cite{mr}.

The latter is based on a functional renormalization group equation (FRGE) which encodes a kind of (continuous) Wilsonian RG flow on the space of diffeomorphism invariant action functionals. These actions naturally depend on the expectation value $g_{\m\n}\=\langle \g_{\m\n} \rangle$ of  the quantum metric $\gamma_{\m\n}$. In addition, the coarse graining operation requires a background structure, which can be used to define volumes over which the quantum fluctuations are averaged. This structure is conveniently provided by the background field method \cite{back} which also ensures the background covariance of the approach. Here the quantum metric is split according to
\be\label{backsplit}
\gamma_{\m\n} = \gb_{\m\n} + h_{\m\n}
\ee
where $\gb_{\m\n}$ is a fixed, but unspecified, background metric and $ h_{\m\n}$ are the quantum fluctuations around this background which are not necessarily small. This allows the formal construction of the gauge-fixed (Euclidean) gravitational path integral
\be\label{PI}
\int \cD h \cD C^\mu \cD \Cb_\mu \exp\{ - S[\gb+h] - S^{\rm gf}[h; \gb] - S^{\rm ghost}[h, C, \Cb; \gb] - \Delta_k S[h, C, \Cb; \gb] \} \, .
\ee
Here $S[\gb+h]$ is a generic action, which depends on $\gamma_{\m\n}$ only, while the background gauge fixing $S^{\rm gf}[h; \gb]$ and ghost contribution $S^{\rm ghost}[h, C, \Cb; \gb]$
 contain $\gb_{\m\n}$ and  $h_{\m\n}$ in such a way that they do not combine into a full $\gamma_{\m\n}$. 
They have an ``extra $\gb_{\m\n}$-dependence'' and are not invariant under split-symmetry $\delta h_{\m\n}=\epsilon_{\m\n}$, $\delta \gb_{\m\n}=-\epsilon_{\m\n}$
which is respected by the combination \eqref{backsplit}. 

The key ingredient in the construction of the FRGE is the coarse graining term
$\Delta_k S[h,C,\Cb; \gb]$. It is quadratic in the fluctuation field $h$, $
\int d^dx \sqrt{\gb} \, h_{\m\n} \cR^{\m\n\rho\sigma}_k(-\Db^2) h_{\rho\sigma}\,$,
plus a  similar term  for the ghosts. 
The kernel $\cR^{\m\n\rho\sigma}_k(p^2)$ provides a $k$-dependent mass term which separates the fluctuations into high momentum modes $p^2 \gg k^2$ and low momentum modes $p^2 \ll k^2$ with respect to the scale set by the covariant Laplacian of the background metric. 
The profile  of  $\cR^{\m\n\rho\sigma}_k(p^2)$ ensures that the high momentum modes are integrated out unsuppressed while the contribution of the low momentum modes to the path integral is suppressed by the $k$-dependent  mass term. Varying $k$ then naturally realizes Wilson's idea of coarse graining by integrating out the quantum fluctuations shell by shell.

Taking the formal $k$-derivative, eq.\ \eqref{PI} provides the starting point for the construction of the functional renormalization group equation for the effective average action $\Gamma_k$  \cite{avact,ym}.(See \cite{avactrev} for reviews.) For gravity this flow equation takes the form \cite{mr}  
\be\label{FRGE}
\p_t \Gamma_k[\hb, \xi, \bar{\xi}; \gb] = \half {\rm STr} \left[ \left( \Gamma_k^{(2)} + \cR_k \right)^{-1} \, \p_t \cR_k  \right]\, .
\ee
Here $t = \log(k/k_0)$, {\rm STr} is a  functional supertrace which includes a minus sign for the ghosts $\xi\=\langle C\rangle, \bar{\xi}\=\langle \Cb\rangle$, $\cR_k$ is the matrix valued (in field space) IR cutoff introduced above, and $\Gamma_k^{(2)}$ is the second variation of $\Gamma_k$ with respect to the {\it fluctuation fields}. Notably, $\Gamma_k[\hb, \xi, \bar{\xi}; \gb]$ depends on {\it two} metrics, the background metric $\gb_{\m\n}$ and the expectation value  field    
\be
g_{\a\b}  \= \langle \gamma_{\a\b} \rangle = \gb_{\a\b} + \bar{h}_{\a\b}  \, , \qquad  \bar{h}_{\a\b} \= \langle h_{\a\b} \rangle \, .
\ee
The explicit dependence on the {\it two} metrics is essential for being able to write down the exact flow equation \eqref{FRGE}, as the Hessian $\Gamma_k^{(2)}$ is the variation of $\Gamma_k$ with respect to the fluctuation fields {\it at fixed} $\gb_{\m\n}$. In this sense, $\Gamma_k$ and its flow is of an intrinsically {\it bimetric} nature. In particular, the construction of $\Gamma_k$ involves the terms $\Delta_k S$ and $S^{\rm gf} + S^{\rm ghost}$ where the $\gb_{\m\n}$-dependence does not combine with $\hb_{\m\nu}$ into the full averaged metric $g_{\m\n}$. These terms therefore provide a source for the {\it extra background field dependence} of $\Gamma_k$. To stress this point, it may be illustrative to write
\be
\Gamma_k[g, \gb, \xi, \bar{\xi}] \equiv \Gamma_k[\hb = g - \gb, \xi, \bar{\xi}; \gb] \, , 
\ee
where $\Gamma_k$ now depends on two full fledged metrics, $g_{\m\n}$ and $\gb_{\m\n}$.

One way to extract physics information from the FRGE is by applying perturbation theory \cite{codello,robertoannph,Codello:2010mj}. 
The main virtue of the flow equation is, however, that its use is not limited to perturbation theory and can also be employed to obtain  non-perturbative information. Here the most common approximation scheme consists of truncating the space of functionals $\Gamma_k$ to a finite-dimensional subspace and projecting the flow equation onto this subspace. 
Studying the gravitational RG flow, within these truncations, the most exciting result obtained to date is a
substantial body of evidence \cite{mr},\cite{codello}-\cite{livrev} in support of Weinberg's asymptotic safety scenario for gravity \cite{wein, wein-kall, wein-bern}.
All truncations of the FRGE have displayed a non-Gaussian fixed point (NGFP) of the gravitational RG flow and there is also mounting evidence \cite{codello,robertoannph,frankmach,BMS} that its number of relevant couplings is actually finite. This fixed point may thus provide a fundamental and predictive UV completion of gravity within Wilson's generalized framework of renormalization. (See \cite{entropy}-\cite{wein-infl} for cosmological applications of this framework.)

While already impressive, a serious caveat in this body of evidence is that all computations carried out to date are essentially ``single-metric'' and do not properly reflect the bimetric nature of the flow equation. Typically the ansatz made for $\Gamma_k$ falls into the class
\be
\Gamma_k[\hb, \xi, \bar{\xi}; \gb] = \bar{\Gamma}_k[g] + \widehat{\Gamma}_k[\hb; \gb] + S^{\rm gf} + S^{\rm ghost} \, ,
\ee
where $\bar{\Gamma}_k[g]$ are interaction monomials built from the expectation value metric only. The split-symmetry violating interactions built from both $\hb, \gb$ are encoded in $ \widehat{\Gamma}_k[\hb; \gb] $ which, by construction, vanishes for $\hb =0$. Finally, $S^{\rm gf}$ and $S^{\rm ghost}$ are taken as the classical gauge-fixing and ghost terms. The single-metric computations then proceed by taking the second variation of $\Gamma_k$ with respect to the fluctuation fields and setting $g = \gb$ afterwards. This suffices to extract the running of the coupling constants contained in $\bar{\Gamma}_k[g]$.

The potentially problematic feature of these computations is that the $\beta$-functions encoding the running of the coupling constants multiplying interactions built from the 
``genuine''  metric $g_{\m\n}$ are tainted by contributions originating from pure background terms. A single-metric truncation does not distinguish between the 
running of, say, 
\be\label{biEH}
I_1 = \frac{1}{16 \pi G_k^{\rm A}} \int d^dx \sqrt{g} \left[ - R(g) + 2 \Lambda_k^{\rm A} \right] \, ,  \; \;  \mbox{and} \qquad \bar{I}_1 = \frac{1}{16 \pi G_k^{\rm B}} \int d^dx \sqrt{\gb} \left[ - R(\gb) + 2 \Lambda_k^{\rm B} \right] \, .
\ee
It determines the running of a linear combination of the couplings $G_k^{\rm A},\Lambda_k^{\rm A} $ and their  background analogs  $G_k^{\rm B},\Lambda_k^{\rm B} $ only.

The bimetric nature of the gravitational average action has been appreciated only very recently, by carrying out preliminary studies in conformally reduced gravity \cite{elisa2}, and studying the bimetric terms induced by quantum effects in the matter sector \cite{MRS1}. Currently there are no results on full-fledged gravity available. There are, however, several good reasons why disentangling between the $g_{\m\n}$ and $\gb_{\m\n}$ contributions is of central importance.
Firstly, the single-metric computations do not account for the bimetric nature of the gauge-, ghost-, and cutoff-terms which inject an extra $\gb$-dependence into the 
path integral. 
They are sources of  split-symmetry breaking action monomials, which will
 inevitably be ``switched on'' along the RG flow, leading to new
  interactions which are either constructed from the background metric only or a mixture of background and expectation value metric.
The preliminary results obtained in \cite{elisa2} and \cite{MRS1} suggest that
 disentangling these interactions may lead to a significant alteration of the results obtained in the single-metric case. In particular, separating the $g$ and $\gb$
pieces in $\G_k $ may destroy the NGFP underlying  Weinberg's Asymptotic Safety idea.
Secondly, identifying $g = \gb$ does not probe the direction of the IR cutoff $\Delta_kS$ in theory space  
which may give rise to a important contribution to the RG flow in the UV. Thirdly, 
 employing the background field method, the counterterms found in perturbation theory are constructed solely from the background fields \cite{WeinIIbook},
 so that 
isolating their effect requires careful distinction between the  $g_{\m\n}$ and $\gb_{\m\n}$ field monomials.
 
Based on this motivation, our work initiates the first study of the full-fledged gravitational RG flow in a fully bimetric setting.
Concretely, we study the  RG flow in the bimetric Einstein-Hilbert truncation which distinguishes the four action monomials in eq.\ \eqref{biEH},
i.e. $\sqrt{g},\, \sqrt{\gb},\, \sqrt{g}\,R,\, \sqrt{\gb}\,\Rb$, respectively with $\Rb\=R(\gb)$.
As our central result, we show that the non-Gaussian fixed point known from the single-metric computations also appears in the bimetric case. 
Subsequently, we supplement the bimetric Einstein-Hilbert truncation by a prototypical one-parameter family of  mixed action monomials built from 
both the ``genuine'' and background metric. 
Including the extra interactions induces a split of the known NGFP in a UV and an IR fixed point, which can be connected by a complete RG trajectory.  

The rest of the paper is organized as follows. In Section \ref{sect:2} we describe the details of the setup and state our main new result: the $\beta$-functions of the double-Einstein-Hilbert truncation in four dimensions. The properties of these $\beta$-functions are analyzed in Section \ref{sect:4} and we discuss our findings in Section \ref{sect:5}. A brief summary of the heat-kernel techniques employed in the paper and the rather lengthy $\beta$-functions for the bimetric Einstein-Hilbert truncation valid for any spacetime dimension $d$ are relegated to the Appendices \ref{App:A} and \ref{App:B}, respectively.

\section{$\beta$-functions of the double-Einstein-Hilbert truncation}
\label{sect:2}
In this section we derive the $\beta$-functions of the double-Einstein-Hilbert truncation.
Besides the Einstein-Hilbert action constructed from $g_{\m\n}$, known from previous single-metric
truncations, the corresponding truncation ansatz also encompasses a Einstein-Hilbert
action constructed from the background metric $\gb_{\m\n}$ and a simple class
of interaction monomials including both $g_{\m\n}$ and $\gb_{\m\n}$. This prototypical setup accounts for
the bimetric nature of the FRGE \eqref{FRGE}, for the first time disentangling the quantum
gravity effects in $g_{\m\n}$ and $\gb_{\m\n}$ in a full gravity computation.\footnote{For a related analysis 
in the framework of conformally reduced  and matter induced gravity, see ref.  \cite{elisa2} and \cite{MRS1}, respectively.}
\subsection{The truncation ansatz}
\label{sect:2.1}
Our ansatz for the double-Einstein-Hilbert truncation takes the form
\begin{equation}\label{EAAA}
\Gamma_k[g,\Cb,C,\gb] = \Gamma^{\rm met}_k[g, \gb] + S^{\rm gf}[h; \gb] + S^{\rm ghost}[g,\Cb,C,\gb] \, ,
\end{equation}
where $\Gamma^{\rm met}_k[g, \gb]$ is the metric part of the effective action (built from both $g$ and $\gb$) which we supplement
 by the classical gauge-fixing and ghost action $S^{\rm gf}$ and $S^{\rm ghost}$, respectively. Explicitly, we consider the following
 one-parameter class of gravitational actions
 \be\label{gr:ansatz}
 \begin{split}
\Gamma^{\rm met}_k[g, \gb] = & - \frac{1}{16 \pi G_k^{\rm A}}\int d^dx \sqrt{g} \left[ R -2 \Lambda_k^{\rm A} \right]
- \frac{1}{16 \pi G_k^{\rm B}} \int d^dx \sqrt{\gb} \left[  \Rb -2 \Lambda_k^{\rm B} \right] \\
& -\frac{M_k}{8 \pi G_k^{\rm A}}\int d^dx\sqrt{g} \left(\frac{\sqrt{\gb}}{\sqrt{g}}   \right)^{n} \, . 
\end{split}
\ee
Here the unbared (bared) quantities are constructed from the expectation value metric $g_{\a\b}$ (background metric $\gb_{\a\b}$). Furthermore, $G_k$ and $\Lambda_k$ 
denote the Newton's constants and cosmological constants, with the superscript $A$ and $B$
indicating that the corresponding interaction term is constructed from $g_{\a\b}$ and $\gb_{\a\b}$ respectively.
The form of the bimetric term appearing in the last line is motivated by the structure of the flow equation encountered in \cite{MRS1},
where it is precisely the ratio $\sqrt{g}/\sqrt{\gb}$ that naturally appears on its right-hand-side. In the following,
we will consider integer exponents  $n \ge 2 $ only. The terms $n=0$, $n=1$ give rise to the monomials
multiplying $ \Lambda_k^{\rm A}$ or $ \Lambda_k^{\rm B}$, respectively, and are already included in the Einstein-Hilbert actions constructed 
from the ``genuine'' and background metric. 

In the sequel, we will work with the geometric gauge-fixing, setting
 \begin{equation}\label{Sgf}
S^{\rm gf} = \frac{1}{2 \alpha} \int d^dx \sqrt{\gb} \gb^{\mu\nu} F_{\mu} F_{\nu} \, , \qquad
F_\mu = \Db^\nu h_{\mu\nu} - \tfrac{1}{d} \Db_\mu h \, , 
\end{equation}
and subsequently taking the Landau gauge limit  $\alpha \rightarrow 0$. As it was shown in \cite{codello,frankmach}, this gauge choice
is perfectly adapted to the transverse-traceless decomposition utilized in Section \ref{sect:2.3} below, where it leads to significant
simplifications. The ghost action exponentiating the resulting Faddeev-Popov determinant takes the form
\begin{equation}
S^{\rm ghost} = - \sqrt{2} \int d^dx \sqrt{\gb} \Cb_\mu \cM^\mu{}_\nu \, C^\nu \, ,  
\end{equation}
with 
\begin{equation}
\cM^\mu{}_\nu  = \gb^{\mu\rho} \gb^{\sigma \lambda} \Db_\lambda (g_{\rho \nu} D_\sigma + g_{\sigma \nu} D_{\rho}) - \tfrac{2}{d}\, \gb^{\rho \sigma} \gb^{\mu\lambda} \Db_\lambda g_{\sigma \nu} D_\rho \, . 
\end{equation}
This completes the specification of our truncation ansatz.

\subsection{The conformal projection technique}
\label{sect:2.2}
%
Our next task is to project the gravitational RG flow onto the subspace spanned by \eqref{EAAA}, so that we can compute the $\beta$-functions for the $k$-dependent couplings contained in $\Gamma_k^{\rm met}$.
Obviously, this cannot be achieved by evaluating the flow equation setting $g_{\a\b} = \gb_{\a\b}$, which underlies the single-metric computations. 
Instead, we will resort to the conformal projection technique introduced in \cite{MRS1} 
which identifies $g_{\m\n}$ and $\gb_{\m\n}$ up to a constant conformal factor:
\begin{equation}\label{background}
g_{\mu\nu} = (1 + \eps)^\nu \, \gb_{\mu\nu} \, , \qquad \nu = \tfrac{2}{d-2} \, , \qquad \ell^2 \equiv 1+\epsilon \, .
\end{equation}
Substituting this identification into \eqref{gr:ansatz} and performing a double-expansion in $\epsilon$ and the background curvature $\Rb$ 
yields
\be\label{LHSflow}
\begin{split}
\left. \Gamma^{\rm met}_k[g, \gb] \right|_{g = (1+\epsilon)^\nu \gb} = & \, \frac{1}{8\pi} \int d^dx  \sqrt{\gb} \, \bigg\{
\left[\tfrac{ \Lambda_k^{\rm B}}{ G_k^{\rm B}}+  \tfrac{ \Lambda_k^{\rm A}}{ G_k^{\rm A}}- \tfrac{M_k}{ G_k^{\rm A}}\right]
+\tfrac{d}{(d-2)} \left[\tfrac{ \Lambda_k^{\rm A}}{ G_k^{\rm A}}- (1-n)\tfrac{M_k}{ G_k^{\rm A}}\right] \eps \\
& \qquad \qquad \qquad \quad +  \tfrac{d}{(d-2)^2} \left[ \tfrac{ \Lambda_k^{\rm A}}{ G_k^{\rm A}} -\tfrac{(1-n)(2-nd)}{2}\tfrac{M_k}{ G_k^{\rm A}} \right] \eps^2
\bigg\} \, \\ 
& \, - \frac{1}{16\pi} \int d^dx  \sqrt{\gb} \Rb \, \bigg\{ \left[ \tfrac{1}{ G_k^{\rm A}} + \tfrac{1}{G_k^B} \right] + \tfrac{1}{ G_k^{\rm A}} \, \epsilon \bigg\} + \cdots \, ,  
\end{split}
\ee
where the dots indicate higher powers in the $\eps, \Rb$ expansion. Plugging this expansion into \eqref{FRGE}, the left-hand-side of the equation
indicates that the running of the coupling constants contained in the ansatz \eqref{gr:ansatz} is captured by the coefficients 
\begin{equation}\label{dexp}
\begin{split}
\Rb^0: & \qquad \eps^0 \, , \qquad \eps^1 \, , \qquad \eps^2 \, , \\
\Rb^1: & \qquad \eps^0 \, , \qquad \eps^1 \, ,
\end{split}
\end{equation}
of this double expansion.
Thus, by extracting the corresponding contributions from the right-hand-side of the flow equation, we are able to disentangle the running
of $G_k^{\rm A},\, \Lambda_k^{\rm A} $ and   $G_k^{\rm B},\, \Lambda_k^{\rm B} $ together with $M_k$.

At this stage, we feel obliged to add the following word of caution. While the conformal projection technique employed here is capable of 
distinguishing between the running coupling constants associated with  monomials built from $g_{\mu\nu}$, the background
metric $\gb_{\mu\nu}$ or a mixture of the two, it has only limited power for resolving different tensorial structures. As an illustrative example,
we consider the following three mixed terms in $d=4$:
\be
 \int d^4x \, (\sqrt{g} \sqrt{\gb})^{1/2} \, , \qquad 
 \int d^4x \sqrt{g} \, (\gb_{\m\n} g^{\m\n}) \, , \qquad
 \int d^4x \sqrt{\gb} \, (\gb^{\m\n} g_{\m\n}) \, . 
\ee
Under the conformal identification \eqref{background} all three invariants are projected onto the same  structure, $ \ell^2 \int d^4x \sqrt{\gb}$,
 and are thus indistinguishable.\footnote{The situation is completely analogous to the projection of the three $R^2$-couplings on a spherical background \cite{codello,oliver2,oliver3,oliver4,BMS}, which also determines the $\beta$-functions for one particular linear combination of the three couplings only.} Resolving this ambiguity will require a much more sophisticated computational technique, like the $h_{\a\b}$-expansion advocated in \cite{MRS1}. Owed to the increased technical complexity of the $h_{\a\b}$-expansion, however, we will refrain from resolving this ambiguity and resort to the conformal projection scheme in the sequel. In any case, we expect that the latter is sufficiently elaborate to give some first insights into the properties of the gravitational RG flow taking the bimetric nature of the flow equation into account.
%
\subsection{Hessian $\Gamma^{(2)}_k$, cutoff implementation, and the flow equation}
\label{sect:2.3}
%
In the next step, it is convenient to first compute the quadratic forms arising at second order in the $h_{\a\b}$-expansion of $\Gamma_k^{\rm met}$. These forms will considerably simplify the computation of the Hessian $\Gamma_k^{(2)}$ later on. We start by constructing the Taylor series of $\Gamma_k^{\rm met}$ around the background \eqref{background}:
\begin{equation}
\Gamma_k^{\rm met}[g, \gb] =  \Gamma_k^{\rm met}[\gb, \gb] + \delta \Gamma_k^{\rm met}[g, \gb]|_{g = \ell^{2 \nu} \gb} + \half \delta^2 \Gamma_k^{\rm met}[g, \gb]|_{g = \ell^{2 \nu} \gb} + \ldots \, . 
\end{equation}
To simplify the notation it is useful to abbreviate the interaction monomials in $\Gamma_k^{\rm met}$ by
\be
I_1 = \int d^dx \sqrt{g} R \, , \qquad I_2 = \int d^dx\sqrt{g}\left(\frac{ \sqrt{\gb}}{\sqrt{g}}\right)^{n} \, .
\ee
The interaction term multiplying $ \Lambda_k^{\rm A}$ and $\Lambda_k^{\rm B}$ are special cases of $I_2$, corresponding to $n=0,1$.

Expanding the curvature invariants up to second order in $h_{\mu\nu}$ and setting $g_{\m\n}=l^{2\nu}\,\gb_{\m\n}$ afterwards, the second variations become
\begin{equation}\label{2var}
\begin{split}
\delta^2 I_1 = & \, 
\int d^dx \sqrt{\gb} \, \ell^{2\alpha_1} \, 
\Big\{
 \tfrac{1}{2} h \left[ \Dz + \tfrac{d^2-5d+8}{2d(d-1)} \Rb \right] h 
- \tfrac{1}{2} h_{\mu\nu} \left[ \Dz + \tfrac{d^2-3d+4}{d(d-1)} \Rb \right] h^{\mu\nu} \\ & \qquad \qquad \qquad \quad
+ h \Db^\mu \Db^\nu h_{\mu\nu} + (\Db_\mu h^{\mu\nu}) (\Db^\alpha h_{\alpha \nu})  
\Big\}
\, , 
\\
\delta^2 I_2 = & \, 
 \tfrac{(1-n)}{2} \int d^dx \sqrt{\gb} \, \ell^{2 \alpha_2} \, \left\{ \tfrac{(1-n)}{2} \, h^2 - h_{\mu\nu} h^{\mu\nu} \right\} \, . 
\end{split}
\end{equation}
 Here $h \= \gb^{\m\n} h_{\m\n}$, $\Dz \= - \Db^2$, and we have freely integrated by parts. All indices are raised and lowered with the background metric.
Furthermore, we have specified the background metric as the one of the $d$-dimensional sphere, satisfying
\be\label{backsphere}
\Rb_{\mu\nu} = \frac{1}{d} \Rb g_{\mu\nu} \, , \qquad \Rb_{\mu\nu\rho\sigma} =  \frac{\Rb}{d(d-1)} \left( \gb_{\mu\rho} \, \gb_{\nu\sigma} - \gb_{\mu\sigma} \, \gb_{\nu\rho} \right) \, , 
\ee
which suffices to keep track of the expansion in the background scalar curvature terms, cf.\ eq.\ \eqref{dexp}. All terms are 
generalized homogeneous in the conformal factor $\ell^2$ and scale with exponents
\be\label{alphaeq}
\alpha_0 = \half \left( d-4 \right) \nu \, , \qquad \alpha_1 = \half \left(d-6 \right) \nu \, , \qquad \alpha_2 = \half \left( d(1-n)-4 \right) \nu \, ,
\ee
for the cosmological constant, $I_1$, and $I_2$, respectively.

In order to diagonalize the Hessian $\Gamma^{(2)}_k$ we implement the transverse-traceless (TT)-decomposition \cite{YorkTT} of the fluctuations fields with 
respect to the spherical background, according to
\be\label{TTdec}
h_{\mu\nu} = h_{\mu\nu}^{\rm T} + \Db_\mu \xi_\nu + \Db_\nu \xi_\mu + \Db_\mu \Db_\nu \sigma + \frac{1}{d} \gb_{\mu\nu} (-\Db^2 \sigma +h ) \, , 
\ee
for the metric fluctuations and
\be\label{Tdec}
\bar{C}_\mu = \bar{C}^{\rm T}_\mu + \Db_\mu \bar{\eta} \, , \qquad C_\mu = C_\mu^{\rm T} + \Db_\mu \eta \, ,  
\ee
for the ghost fields, respectively. The component fields are subject to the (differential) constraints
\be
\begin{split}
\gb^{\m\n} h_{\mu\nu}^{\rm T} = 0 \, , \quad \Db^\mu h_{\mu\nu}^{\rm T} = 0 \, , \quad \Db^\mu \xi_\mu = 0 \,  , \quad
\Db^\mu \bar{C}_\mu^{\rm T} = 0 \, , \quad \Db^\mu C_\mu^{\rm T} = 0 \, . 
\end{split}
\ee
The resulting Jacobian determinants resulting from the TT-decomposition are exponentiated by introducing suitable auxiliary fields along the lines of the Faddeev-Popov trick.
For this purpose, we introduce the transverse vector ghosts $\bar{c}^{{\rm T} \mu},\, c_\mu^{\rm T}$, the transverse vector $b_\mu^{\rm T}$, the scalar ghosts $\bar{c},c$, the real scalar $b$, and a complex scalar $\bar{s},s$, which enter into the auxiliary action (see \cite{codello,frankmach} for more details). On the spherical  background 
\eqref{backsphere} it reads: 
\be\label{Saux}
\begin{split}
S^{\rm aux} = \int d^dx \sqrt{\gb} \Big\{ & 
 \bar{c}_\mu^{\rm T} \left[ \Dz - \tfrac{1}{d} \Rb \right] c^{{\rm T}\mu} + \tfrac{d-1}{d} \, \bar{c} \left[ \Dz - \tfrac{1}{d-1} \Rb \right] \Dz \, c \\ &
+ b_\mu^{\rm T} \left[ \Dz - \tfrac{1}{d} \Rb \right] b^{{\rm T}\mu} + \tfrac{d-1}{d} \, b \left[ \Dz - \tfrac{1}{d-1} \Rb \right] \Dz \, b + \bar{s} \Dz s
\Big\} \, .
\end{split}
\ee

Substituting the decomposition \eqref{TTdec}, it is now straightforward to obtain the  component field representation of \eqref{2var}
\be\label{metdec}
\begin{split}
\delta^2 I_1 = & \int d^dx \sqrt{\gb} \, \ell^{2\alpha_1}  \, \Big\{
\tfrac{(d-2)(d-1)}{2d^2} h \left[\Dz + C_S  \Rb \right] h 
 - \tfrac{1}{2} h^{\rm T}_{\mu\nu} \left[ \Dz + C_T \Rb \right] h^{{\rm T} \mu\nu}
- \tfrac{d-2}{d} \xi^\mu \Rb \left[ \Dz - \tfrac{1}{d} \Rb \right] \xi_\mu \\
& \qquad \qquad  
- \tfrac{1}{d} h \left[ (d-1) \Dz - \Rb \right] \Dz \sigma 
+ \sigma \left[ \tfrac{(d-2)(d-1)}{2d^2} \Dz^2 - \tfrac{d-2}{2d} \Rb \Dz + \tfrac{d^2-3d+3}{d^2(d-1)} \Rb^2 \right] \Dz \sigma
\Big\} \, , \\
\delta^2 I_2 = & \tfrac{1-n}{2} \int d^dx \sqrt{\gb} \, \ell^{2\a_2}  \, \Big\{ \tfrac{(1-n)d-2}{2d} h^2 - h_{\mu\nu}^{\rm T} h^{\rm T{\mu\nu}} - 2 \xi^\mu \left[ \Dz -  \tfrac{1}{d} \Rb \right] \xi_\mu 
- \tfrac{1}{d} \sigma \left[ (d-1) \Dz - \Rb \right] \Dz  \sigma  \Big\} \, ,
\end{split}
\ee
where we abbreviated
\be\label{cdef}
C_T \= \frac{d^2-3d+4}{d(d-1)} \, , \qquad C_S \= \frac{d-4}{2(d-1)} \, . 
\ee
Analogously, the quadratic form arising from the gauge-fixing term reads
\be\label{Sgfdec}
\delta^2 S^{\rm gf} = \frac{1}{2 \alpha} \int d^dx \sqrt{\gb} \left\{
\xi_\mu \left[ \Dz  - \tfrac{1}{d} \Rb \right]^2 \xi^\mu + \tfrac{1}{d^{2}} \sigma \left[ (d-1) \Dz - \Rb \right]^2 \Dz  \sigma
\right\} \, ,
\ee
while the ghost action gives
\be\label{Sghdec}
\delta^2 S^{\rm ghost} = \sqrt{2} \int d^dx \sqrt{\gb} \, \ell^{2 \nu} \, \left\{
\Cb^{\rm T}_\mu \left[ \Dz - \tfrac{1}{d} \Rb \right] C^{\mu {\rm T}}
+ \tfrac{2}{d} \, \bar{\eta} \left[ (d-1) \Dz - \Rb \right] \Dz \,  \eta 
\right\} \, .
\ee

Reinstalling the $k$-dependent coupling constants multiplying the action monomials,
it is then straightforward to compute the Hessian
\be
\left[ \Gamma^{(2)}_k \right]^{ij}(x,y) =  (-1)^{[j]} \frac{1}{\sqrt{\gb(x)}} \frac{1}{\sqrt{\gb(y)}} \frac{\delta^2 \Gamma_k}{\delta \varphi_i(x) \delta \varphi_j(y) } 
\ee
where $\varphi\=(\varphi_i) \= \left( h^{\rm T}_{\a\b}, h, \xi_\mu, \sigma; \bar{C}^{\rm T}_\mu , C_\mu^{\rm T} , \bar{\eta} , \eta;  \, \bar{c}^{\rm T}_\mu, c_\mu^{\rm T}, b_\mu^{\rm T}, \bar{c}, c, b, \bar{s}, s\right)$ is the multiplet of all fluctuation fields,  and $(-1)^{[j]}$ takes values 0 or 1 for $\varphi_j$  Grassmann-even or odd,
respectively. The matrix elements of $\left[ \Gamma^{(2)}_k \right]^{ij}$ in field space are then summarized in the second column of Table \ref{t.1}. In order to uniformize the expressions  in the gravitational sector, we introduced the
$d, n$-dependent constants
\be\label{csconst}
\tilde{c}_0 \=  \frac{d}{(d-1)(d-2)} \, \Big(d(1-n) - 2\Big) \, (1-n) \, , \qquad \tilde{c}_{\rm 2T} \= 2 (1-n) \, .
\ee
\begin{table}[t]
\begin{center}
\begin{small}
\begin{tabular}{|c|c|c|}
\hline
\bf{Fields} & \bf{Hessian} $\Gamma_k^{(2)}$ & \bf{Kernel} $\cR_k$ \\ \hline
\parbox[0pt][2em][l]{0cm}{} $ \!\! \! h^{\rm T} h^{\rm T}$ & 
$\tfrac{1}{32\pi G_k^{\rm A}} \left[ \ell^{2\alpha_1} \Dz + \ell^{2\alpha_1} C_T \Rb  - 2\Lambda_k^{\rm A} \ell^{2 \alpha_0} +  \tilde{c}_{\rm 2T}\, M_k \ell^{2 \alpha_2} \right]$ &
$\tfrac{1}{32\pi G_k^{\rm A}} R_k$ \\
\parbox[0pt][2em][l]{0cm}{} $\!\! \! hh$ & 
$ -\tfrac{(d-2)(d-1)}{ 32 \pi d^2 G_k^{\rm A}} \left[ \ell^{2\alpha_1} \Dz  +  \ell^{2\alpha_1} C_S  \Rb  - \tfrac{d}{d-1}\Lambda_k^{\rm A} \, \ell^{2\alpha_0} + \tilde{c}_0 \, M_k \, \ell^{2\alpha_2} \right]$ & 
$-\tfrac{(d-2)(d-1)}{ 32 \pi d^2 G_k^{\rm A}} \,  R_k $\\ \hline
\parbox[0pt][2em][l]{0cm}{} $\!\! \! \xi \xi$ &
$\tfrac{1}{\alpha} \left[ \Dz - \tfrac{1}{d}  \Rb \right]^2 $ & 
$ \tfrac{1}{\alpha} \left[ \Pb_k^2 - \Dz^2 - \tfrac{2}{d} \Rb R_k \right] $\\ 
\parbox[0pt][2em][l]{0cm}{} $\!\! \! \sigma \sigma$ & $\tfrac{(d-1)^2}{\alpha d^2} \left[ \Dz - \tfrac{1}{d-1} \Rb \right]^2 \Dz$ & $\cR_k^{\sigma\sigma} $
 \\ 
\parbox[0pt][2em][l]{0cm}{} $\!\! \! \Cb^{\rm T}_\mu C^{\rm T}_\mu$ & $\sqrt{2} \, \ell^{2\nu} \left[ \Dz - \tfrac{1}{d} \Rb \right] $ &
$\sqrt{2} R_k $ \\
\parbox[0pt][2em][l]{0cm}{} $\!\! \! \bar{\eta} \eta$ & $2 \sqrt{2} \, \ell^{2\nu}  \tfrac{d-1}{d} \left[ \Dz - \tfrac{1}{d-1} \Rb \right] \Dz $ &
 $ \tfrac{2 \sqrt{2} (d-1)}{d} \left[ \Pb_k^2-\Dz^2 - \tfrac{\Rb}{d-1}  R_k \right] $ \\ \hline
\parbox[0pt][2em][l]{0cm}{} $\!\! \! \bar{c}^{{\rm T} \mu} c_\mu^{\rm T}$ & $ \Dz - \frac{1}{d} \Rb $ &
$ R_k $ \\
\parbox[0pt][2em][l]{0cm}{} $\!\! \! b^{{\rm T} \mu} b_\mu^{\rm T}$ & $ \Dz - \frac{1}{d} \Rb $ & 
$R_k$ \\
\parbox[0pt][2em][l]{0cm}{} $\!\! \! \bar{c} c$ & $\tfrac{d-1}{d} \left[ \Dz - \tfrac{1}{d-1} \Rb \right] \Dz $ 
& $\tfrac{d-1}{d}  \left[ \bar{P}_k^2 - \Dz^2 - \tfrac{1}{d-1} \Rb R_k \right] $ \\
\parbox[0pt][2em][l]{0cm}{} $\!\! \! b b$ & $\tfrac{d-1}{d}  \left[ \Dz - \tfrac{1}{d-1} \Rb \right] \Dz $ & $\tfrac{d-1}{d}  \left[ \bar{P}_k^2 - \Dz^2 - \tfrac{1}{d-1} \Rb R_k \right] $ \\ 
\parbox[0pt][2em][l]{0cm}{} $\!\! \! \bar{s}{s}$ &  $\Dz$ & $R_k$ \\ \hline
\end{tabular}
\end{small}
\end{center}
\caption{\small{ Matrix elements of  $\Gamma_k^{(2)}$ and the coarse graining kernel to leading order in the gauge-fixing parameter $\alpha$. The horizontal lines separate the contributions from the gravitational, ghost, and auxiliary sector, respectively. The explicit expression for  $\cR_k^{\sigma \sigma}$ is provided in eq.\ \eqref{sscutoff}.}}\label{t.1}
\end{table}
Notably, the entries of $\Gamma_k^{(2)}$ in the  $\xi\xi$ and $\sigma\sigma$ sector contain the contribution from the gauge-fixing term only, and omit the terms originating from $\Gamma_k^{\rm met}$. The latter are subleading in $\alpha$ and can be shown to drop out of the flow equation once the Landau limit $\alpha \rightarrow 0$ is taken. Anticipating this result, Table \ref{t.1} gives only the leading $\alpha$-terms.

The next step in obtaining the $\beta$-functions is the construction of the matrix-valued IR-cutoff operator $\cR_k$. This operator  provides a $k$-dependent mass term for the fluctuation fields which is built from the background metric only. This implies that $\cR_k$ cannot depend on $\epsilon$. The $\epsilon$-dependence of $\Gamma_k^{(2)}$ then enforces a modification of the cutoff schemes used in previous single-metric computations. Focusing on the cutoff of Type I \cite{robertoannph} the rule for determining $\cR_k$ in a single-metric truncation adjusts $\cR_k$ in such a way that all covariant Laplacians are dressed by a $k$-dependent mass-term according to
\be\label{TypeISM}
\Dz \mapsto \Dz + R_k \equiv \bar{P}_k \, . 
\ee
Here $R_k = k^2 R^{(0)}(\Dz/k^2)$ and $R^{(0)}(z)$ is a shape function interpolating monotonously between $R^{(0)}(0) = 1$ and $\lim_{z \rightarrow \infty} R^{(0)}(z) = 0 $. For the bimetric setup of this paper, we generalize this rule in the minimal sense  
\be\label{TypeIBM}
\left. \Dz \right|_{\eps = 0} \mapsto \left. \Dz + R_k \right|_{\eps = 0} \, ,
\ee
i.e. $R_k$ provides a $k$-dependent mass term at zeroth order in the $\epsilon$-expansion. This definition  reduces to the standard Type I cutoff implementation for the single-metric case.

Applying \eqref{TypeIBM} to the matrix entries $\Gamma_k^{(2)}$ then determines the entries of $\cR_k$ uniquely. The result is displayed in the third column of Table \ref{t.1}, with the explicit form of $\cR_k^{\sigma\sigma}$ being
\be\label{sscutoff}
\cR_k^{\sigma\sigma} = \tfrac{(d-1)^2}{\alpha d^2} \left[ \left(\Pb_k - \tfrac{1}{d-1} \Rb\right)^2 \Pb_k - \left(\Dz- \tfrac{1}{d-1}\Rb\right)^2 \Dz \right] \, . 
\ee

With this result, we now have all ingredients for the explicit construction of the operator trace appearing on the right-hand-side of the flow equation resulting from our ansatz.
Utilizing the block diagonal form of $\Gamma_k^{(2)}$ in field-space, this trace decomposes as 
\be\label{flowev}
\left. \p_t \Gamma_k^{\rm met}[g, \gb] \right|_{g = (1+\epsilon)^\nu \gb} = \cS_{\rm 2T} + \cS_{0} + \cS_{\rm gf} + \cS_{\rm aux} \, . 
\ee
Here the left-hand-side is given by the $\p_t$-derivative of the double expansion \eqref{LHSflow}, while $\cS_{\rm 2T}$, $\cS_{0}$, $\cS_{\rm gf}$, and $\cS_{\rm aux}$ are the operator traces constructed from the transverse-traceless $h^{\rm T}_{\m\n}$-fluctuations, the metric scalar $h$, the gauge-fixing sector in the second block, and the auxiliary field contribution given in the third block of Table \ref{t.1}, respectively. By first carrying out a double expansion of the trace-arguments with respect to $\eps, \Rb$, retaining all the terms indicated in \eqref{dexp}, the traces can be evaluated using standard early-time heat-kernel techniques. Since the corresponding computation is rather technical, it has been relegated to Appendix \ref{App:B}, where we also give the explicit expressions for the $d$-dimensional $\beta$-functions. For the rest of the paper we restrict ourselves to the case $d=4$ for which the explicit $\beta$-functions are given in the next subsection.

We close this subsection with a remark on the unphysical exceptional modes arising from working with the TT-decomposition on a spherical background. Performing a spectral decomposition of the component fields $\xi_\mu, \sigma$ in terms of $\Dz$-eigenmodes, one finds that the two lowest scalar eigenmodes (the constant mode and the lowest non-trivial eigenfunction satisfying the conformal Killing equation) and the lowest vector-eigenmode (satisfying the Killing equation) do not contribute to $h_{\m\n}$ and therefore require special care when evaluating the operator traces \eqref{flowev}. One finds that their contribution to the flow equation enters only at $\cO(\Rb^2)$, however, so that this subtlety can safely be disregarded in the present computation.

\subsection{The four-dimensional $\beta$-functions}
\label{sect:2.4}
Based on eq.\ \eqref{flowev}, the $\beta$-functions
\be\label{betadef}
\p_t \mathbf{g}_i(k) = \beta_{i}(\mathbf{g}) \, , \qquad \mathbf{g} = \{ \lambda_k , g_k , m_k , \lambda_k^{\rm B} , g_k^{\rm B} \} \,, 
\ee
arising in the $d$-dimensional double-Einstein-Hilbert truncation are computed in Appendix \ref{App:B}. To simplify our notation
we will set $d=4$ in the following and work with the $4$-dimensional version of the dimensionless coupling constants \eqref{dimless}
\be\label{fourless}
\l_k = k^{-2} \Lambda_k^{\rm A} \, , \quad g_k = k^{2} G_k^{\rm A} \, , \quad m_k = k^{-2} M_k \, , \quad \lambda^{\rm B}_k = k^{-2} \Lambda^{\rm B}_k \, , \quad g^{\rm B}_k = k^{2} G^{\rm B}_k \, .
\ee
Furthermore, we define the anomalous dimensions of the two Newton constants as
\be\label{etaN}
\eta_N = (G_k^{\rm A})^{-1} \p_t G_k^{\rm A} \, , \qquad  \eta_N^{\rm B} = (G_k^{\rm B})^{-1} \p_t G_k^{\rm B} \, . 
\ee

The $\beta$-functions are most conveniently  expressed in terms of the dimensionless threshold functions \eqref{phidef}.
In this context, it turns out to be convenient to introduce a short-hand notation for the $\Delta$- and $\Rb$-independent terms appearing in the square brackets in the first and second line of Table \ref{t.1}. The arguments of the threshold functions are then given by the  
 $\eps = 0$-limit of these terms. 
For the $h^{\rm T}_{\a\b}$ and $h$-contributions they read
\be
w_{\rm 2T} = 2 \Big[ (1-n) m_k - \lambda_k \Big] \, , \qquad w_0 = \tfrac{4}{3} \Big[ (n-1)(2n-1) m_k - \lambda_k \Big] \, .
\ee
Furthermore, we denote the first and second derivative of these terms with respect to $\eps$ by
\be
\begin{array}{ll}
w_{\rm 2T}^\prime = 4n(n-1) \, m_k   \, , 
&w_0^\prime =  - \tfrac{8}{3} n (n-1)(2n-1) \, m_k     \, , \\[1.1ex]
w_{\rm 2T}^{\prime\prime} =-4 n (n-1)  (2n+1) \, m_k   \, , 
& w_{\rm 0}^{\prime\prime} =  \tfrac{8}{3} n (n-1)  (2n-1)(2n+1) \, m_k\, .
\end{array}
\ee
These expressions can also be obtained by setting $d=4$ in \eqref{arg} and \eqref{argder}, respectively.

Using these notations the $\beta$-functions in four spacetime dimensions can be summarized as follows. First we explicitly solve \eqref{etafct} from Appendix B for $\eta_N$. This yields the anomalous dimension

\be\label{eta4}
\eta_N = \frac{g_k\; B_1(\l_k, m_k)}{1- g_k\; B_2(\l_k, m_k)} \, . 
\ee
The $B_1$ and $B_2$ are obtained by splitting $F^{(1,1)} \equiv (16 \pi)^{-1} (B_1 + \eta_N B_2)$ into the terms independent and linear in $\eta_N$:
\be
\begin{split}
B_1 = & \, \frac{1}{6\pi} \Big\{ 25 \;\Phi^2_2(w_{\rm 2T}) + \Phi^2_2(w_{\rm 0}) - 5 w_{\rm 2T}^\prime\; \Phi^2_1(w_{\rm 2T}) - w_0^\prime \;\Phi^2_1(w_{\rm 0}) \\ & \,
\qquad \quad + 40\; w_{\rm 2T}^\prime \;\Phi^3_2(w_{\rm 2T}) - 80\; \Phi^3_3(w_{\rm 2T}) - 3\;\Phi^2_2 + 28\; \Phi^3_3 + 72 \;\Phi^4_4   \Big\} , \\
B_2 = & \, -\frac{1}{12\pi} \Big\{ 25\; \Pt^2_2(w_{\rm 2T}) + \Pt^2_2(w_{\rm 0}) - 5 w_{\rm 2T}^\prime \;\Pt^2_1(w_{\rm 2T}) - w_0^\prime\; \Pt^2_1(w_{\rm 0}) %
\\ & \, \qquad \quad
+ 40 w_{\rm 2T}^\prime\; \Pt^3_2(w_{\rm 2T}) - 80\; \Pt^3_3(w_{\rm 2T}) \Big\} .
\end{split}
\ee
Here the threshold functions $ \Phi^p_n(w)$ and $\Pt^p_n(w)$ without explicit argument are understood to be
evaluated at $w=0$. Utilizing \eqref{eta4} the $\beta$-functions for $g_k, \lambda_k$ and $m_k$ 
can be obtained from equation \eqref{betaphys}:
\begin{subequations}\label{betaphys4}
\begin{align}
\beta_g = & \left[\, 2 + \eta_N \, \right]g_k \, , \\
\beta_{\l} = & (\eta_N - 2) \lambda_k + \frac{\pi }{ n } \, g_k\, \left[ 4 F^{(2,0)} + 2 (2n-1) F^{(1,0)} \right] \, , \\
\beta_{m} = & (\eta_N - 2) m_k +  \, \frac{2 \pi }{ n (1-n)} \,g_k\,  \left[ 2 F^{(2,0)} - F^{(1,0)} \right] \, .
\end{align}
\end{subequations}
The running of the background couplings is governed by
\begin{subequations}\label{betaback4}
\begin{align}
\beta_{g^{\rm B}} = &  \left[ 2 + \eta_N^{\rm B} \right]  \, g^{\rm B}_k \, , \\
\beta_{\lambda^{\rm B}} = & (\eta_N^{\rm B} - 2) \lambda_k^{\rm B} + \frac{\pi  }{(1-n)}\,g_k^{\rm B} \,\left[ 8(1-n) F^{(0,0)} + 4 F^{(2,0)} - 2(3-2n) F^{(1,0)} \right] \, ,
\end{align}
\end{subequations}
where the background anomalous dimension is given by 
\be\label{etab4}
\eta_N^{\rm B} = 16 \pi \left( F^{(0,1)} - F^{(1,1)} \right) g_k^{\rm B} \, . 
\ee
The general result for the functions $F^{(i,j)}(g_k, \lambda_k, m_k)$ is given in Appendix \ref{App:B}. Setting $d=4$ it simplifies to:
\be\label{Ffct4}
\begin{split}
F^{(0,0)}= &  \frac{1}{(4\pi)^{2}}\left[ 5 \, q^1_{2}(w_{\rm 2T}) + q^1_{2}(w_0) - 4 \;\Phi_{2}^1 \right] \, , \\
F^{(0,1)}= & \frac{1}{(4\pi)^{2}} \Big[ \tfrac{5}{6}q^1_{1}(w_{\rm 2T})- \tfrac{10}{3} q^2_{2}(w_{\rm 2T}) +
 \tfrac{1}{6}q^1_{1}(w_{0})
- \tfrac{13}{12} \;\Phi_{2}^2-\tfrac{2}{3}\;\Phi_{1}^1
 \Big] \, , \\
 F^{(1,0)} = & \frac{1}{(4\pi)^{2}} \Big[10 q^2_{3}(w_{\rm 2T}) + 2 q^2_{3}(w_0)
  -5 \, w_{\rm 2T}^{\prime} \, q^2_{2}(w_{\rm 2T}) - w_0^{\prime} \, q^2_{2}(w_0) + 24  \;\Phi^3_{4} + 12  \;\Phi^2_{3} \Big] \, ,\\
 F^{(1,1)} = & \frac{1}{(4\pi)^{2}} \Big[ - 
 \tfrac{1}{6}\,w_0^{\prime}\,  q^2_1(w_0) - \tfrac{5}{6}\, w_{\rm 2T}^{\prime}\, q^2_1(w_{\rm 2T}) +\tfrac{1}{6}  q^2_2(w_0) + 
 \tfrac{25}{6} q^2_2(w_{\rm 2T}) + \tfrac{20}{3}\,  w_{\rm 2T}^{\prime}\,  q^3_2(w_{\rm 2T}) 
\\ &
- \tfrac{40}{3} q^3_3(w_{\rm 2T}) 
-\half \;\Phi^2_2 + \tfrac{14}{3}\;\Phi^3_3 + 12 \Phi^4_4\Big] \, ,\\
F^{(2,0)} = & \frac{1}{(4\pi)^{2}} \Big[ 
30  q^3_{4}(w_{\rm 2T}) + 6 q^3_{4}(w_{\rm 0}) 
-10   q^2_{3}(w_{\rm 2T}) -2 q^2_{3}(w_{\rm 0})
 -20 \, w_{\rm 2T}^{\prime} \,  q^3_{3}(w_{\rm 2T}) -4 w_{\rm 0}^{\prime} \, q^3_{3}(w_{\rm 0}) 
\\ &
+  5 \, (w_{\rm 2T}^{\prime})^2 \,  q^3_{2}(w_{\rm 2T}) + (w_{\rm 0}^{\prime})^2 \, q^3_{2}(w_{\rm 0})  
-  \tfrac{5}{2} w_{\rm 2T}^{\prime\prime} \, q_{2}^2(w_{\rm 2T}) - \half  w_{\rm 0}^{\prime\prime} \, q_{2}^2(w_{\rm 0}) 
 - 36 \;\Phi^3_{4} - 480  \Phi^5_{6}
\Big] \, ,
\end{split}
\ee
The $\beta$-functions \eqref{betaphys4} and \eqref{betaback4} together with \eqref{eta4} and \eqref{etab4} constitute the main result of this section. 
Their properties, in particular the   fixed point structure they give rise to,  will
be investigated in the next section.

\subsection{Bimetric vs.\ single-metric truncations}

At this point it is illustrative to compare the bimetric ansatz \eqref{gr:ansatz} to the single-metric (SM) Einstein-Hilbert truncations studied previously \cite{mr,robertoannph,oliver1,frank1,souma,litimgrav} where 
\be\label{smEH}
\Gamma^{\rm met}_k[g, \gb] =  - \frac{1}{16 \pi G_k^{\rm SM}}\int d^dx \sqrt{g} \left[ R -2\Lambda_k^{\rm SM} \right] \, .
\ee
It is this functional which, for the Einstein-Hilbert case, corresponds to $\bar{\Gamma}_k[g]$ in eq.\ \eqref{gr:ansatz}. It depends on the background only via 
$g_{\m\n}\=\gb_{\m\n}+ \hb_{\m\n}$, that is, it has no {\it extra} background dependence: $\Gamma^{\rm met}_k[g, \gb]=\bar{\Gamma}_k[g]$.

How is the functional \eqref{smEH}
 and its predictions for the RG flow related to those of the bimetric ansatz \eqref{gr:ansatz}? The correct mapping of the bimetric computation onto its single
metric analog consists in simply omitting the second and third integral on the RHS of \eqref{gr:ansatz}. This amounts to setting $1/ G_k^{\rm B}\=0,\,\Lambda_k^{\rm B}\=0,\, 
M_k\=0 $ and identifying
\be\label{ident} 
G_k^{\rm SM}=G_k^{\rm A},\qquad \Lambda_k^{\rm SM}=\Lambda_k^{\rm A}.
\ee

While this mapping scheme is very natural, the reader might argue that there is another one that seems equally plausible. In a certain sense, 
the single metric truncation does not distinguish $g_{\m\n}$ from $\gb_{\m\n}$, so one could be motivated to keep all three integrands in \eqref{gr:ansatz},
but replace $\gb_{\m\n}$ with the expectation value metric   $g_{\m\n}$ everywhere. The result is a single metric functional of the form \eqref{smEH} with 
the couplings
\be\label{identwrong}
\frac{1}{G_k^{\rm SM}}=\frac{1}{ G_k^{\rm A}} + \frac{1}{G_k^{\rm B}}  \, , \qquad  \frac{\Lambda_k^{\rm SM}}{G_k^{\rm SM}}= \frac{ \Lambda_k^{\rm A}}{ G_k^{\rm A}} + 
\frac{ \Lambda_k^{\rm B}}{G_k^{\rm B}} - \frac{ M_k}{ G_k^{\rm A}} \, . 
\ee
We must emphasize that the identification \eqref{identwrong} is actually {\it not} the correct way of relating the new bimetric calculation to their old single metric
counterpart. The reason is that if one proceeds in this way one has to retain the second and third integral of \eqref{gr:ansatz}, with $g=\gb$, also on the
RHS of the flow equation. The identifications \eqref{identwrong} amount to setting $g=\gb$ directly in the ansatz, i.e. {\it prior to computing the Hessian}.
As a result, $\G_k^{(2)}$, in this case, receives contributions also from the second and third integral of \eqref{gr:ansatz}, involving the background couplings $G_k^{\rm B}$,
 $\Lambda_k^{\rm B}$, and $M_k$. Since those contributions are absent in the actual bimetric calculation it is clear that \eqref{identwrong} cannot be correct.

Within our present computational setting 
the $\beta$-functions for the dimensionless single-metric couplings $\{g^{\rm SM},\l^{\rm SM}\}$ are easily
derived noticing that the RHS of the single-metric RG flow equation is given by the zeroth order terms
$ F^{(0,0)}(g^{\rm SM},\l^{\rm SM})$ and $ F^{(0,1)}(g^{\rm SM},\l^{\rm SM})$ in the $\epsilon$-expansion of \eqref{Fexp}. Thus, the equations read
\begin{subequations}\label{sm-betafct}
\begin{align}
\beta_{g^{\rm SM}} & = \left[\, 2 + \eta_N^{\rm SM} \, \right]g^{\rm SM}_k,\label{sm-a}\\
\beta_{\l^{\rm SM}} & =  (\eta_N^{\rm SM} - 2) \lambda_k^{\rm SM} + 8 \pi g_k^{\rm SM} F^{(0,0)}\label{sm-b} \, ,  
\end{align}
\end{subequations}
where 
$\eta_N^{\rm SM}=16\pi g_k^{\rm SM}  F^{(0,1)}$ is the anomalous dimension of $G_k^{\rm SM}$. We shall come back to these $\beta$-functions shortly.
%

\section{RG flow of the double-Einstein-Hilbert  truncation}
\label{sect:4}
We now investigate the RG flow resulting from the $\beta$-functions \eqref{betaphys4} and \eqref{betaback4}. The system without mixed term, $m_k = 0$, is analyzed in Subsection \ref{sect:4.1} while the properties of the full system are discussed in Subsection \ref{sect:4.2}.
%
\subsection{The double-Einstein-Hilbert truncation without mixed term}
\label{sect:4.1}

In order to understand the properties of the bimetric RG flow, we first  discuss 
the simpler four-parameter truncation without the mixed term $\propto \left(\sqrt{\gb}/\sqrt{g}\right)^n$.  The corresponding $\beta$-functions can be 
recovered from eqs.\ \eqref{betaphys4} and \eqref{betaback4} which requires some care though. (See the remarks at the end of Appendix B.) In $d=4$, the system
of $\beta$-functions reads
\begin{subequations}\label{4.1}
\begin{align}
\beta_{g} = & \left[\, 2 + \eta_N \, \right]g_k \, ,\label{4.1a} \\
\beta_{\l} = & (\eta_N - 2) \lambda_k + 4 \pi g_k\; F^{(1,0)} \, ,\label{4.1b} \\
\beta_{g^{\rm B}} =&  \left[ 2 + \eta_N^{\rm B} \right]  \, g^{\rm B}_k, \, \label{4.1c}\\
\beta_{\lambda^{\rm B}} = & (\eta_N^{\rm B} - 2) \lambda_k^{\rm B} + 8 \pi g_k^{\rm B} \left[  F^{(0,0)} -\;\half \; F^{(1,0)} \right]\label{4.1d} \, ,
\end{align}
\end{subequations}
where $\eta_N$ and $\eta_N^{\rm B}$ are the anomalous dimensions defined in \eqref{eta4} and \eqref{etab4}, respectively. 
This system is decoupled in the following sense.
Eqs.\ \eqref{4.1a} and \eqref{4.1b} close among themselves and are independent of the background couplings $g^{\rm B}$ and $\lambda^{\rm B}$.
Once a solution of this subsystem is given, we can insert it into  the remaining $\beta$-functions \eqref{4.1c} and \eqref{4.1d}
and solve the resulting two differential equations for $g_k^{\rm B}$ and $ \lambda_k^{\rm B} $.
To find fixed point solutions in this four-parameter system, it is enough to search for fixed points  $(g_*,  \lambda_*)$ of the subsystem \eqref{4.1a} and \eqref{4.1b},
substitute  $(g_*, \lambda_*)$ into  \eqref{4.1c} and \eqref{4.1d},
 and then look for zeros  $(g_*^{\rm B},  \lambda_*^{\rm B})$  of the background $\beta$-functions.

Following this strategy and using the optimized shape function \cite{opt}, a numerical search unveils the following fixed point structure.
The subsystem \eqref{4.1a}-\eqref{4.1b} admits a Gaussian fixed point (GFP) at $(g_*,\lambda_*) = 0$ and a NGFP
with $(g_*,\lambda_*)\neq 0$. Each of these fixed points
gives rise to  a pair of zeros of  the background $\beta$-functions \eqref{4.1c}-\eqref{4.1d}. 
One of them corresponds to a background GFP with  $(g_*^{\rm B},\lambda_*^{\rm B})= 0$, the other to a background NGFP at  $(g_*^{\rm B},\lambda_*^{\rm B})\neq 0$.
The resulting four combinations of fixed points are summarized in  Table \ref{tab:mk=0}.

Having found a fixed point  $\mathbf{g}_{*}$ of the four-parameter system $\mathbf{g}\equiv(\mathbf{g}_i)\equiv (g, \lambda,g^{\rm B},\lambda^{\rm B})$, 
we compute the stability
matrix $B_{ij}=\partial_j\beta_i(\mathbf{g}_*)$ which governs the RG flow linearized around the fixed point:
\begin{equation}\label{stabM}
k\partial_k\; \mathbf{g}_i(k)= B_{ij}\Big(\mathbf{g}_j(k)-\mathbf{g}_{*j}\Big) \, .
\end{equation}
Setting $t=\ln(k)$, the general solution of \eqref{stabM} reads, in the nondegenerate case,
\begin{equation}\label{solstabM}
\mathbf{g}_j(k)=\mathbf{g}_{*j}+ \sum_n r_n e^{i\alpha_n}\;e^{-\theta_n t}\;\mathbf{V}_j^n \, .
\end{equation}
Here $\mathbf{V}^n$ are the right eigenvectors of the stability matrix, with eigenvalues $-\theta_n$, and
$r_ne^{i\alpha_n}\equiv C_n$ are free constants of integration.
They can be complex except when $\theta_n$ happens to be real (then $\alpha_n=0$). Critical exponents with
$\textrm{Re}(\theta_n)>0$ correspond to relevant scaling fields. They grow when $k$ is lowered, i.e. they amount
to UV attractive directions.

Based on the results shown in Table \ref{tab:mk=0}, the stability matrix governing the linearized flow near the NG-NG-FP can be evaluated numerically.
Upon its diagonalization we find  a pair of complex conjugate critical exponents
$\theta_1=\theta_2^*\equiv \theta'+i\theta''$ with
\be
\theta'=\,4.468,\qquad\theta''=\,4.240,
\ee
 together with the background critical exponents 
 \be\label{thetaback}
\theta_4 = 4\quad\textrm{ and  }\quad \theta_5= 2 \, .
\ee
Since all critical exponents are positive,
 the NG-NG-FP is UV-attractive in all directions. While $\theta'$ and $\theta''$ depend on the cutoff chosen, our results for $\theta_4$ and $\theta_5$ are
 universal, i.e.\ cutoff independent. This follows from the special structure of the system \eqref{4.1}, which implies a stability matrix which has a lower triangular form. Consequently, the critical
exponents of the background $\beta$-functions \eqref{4.1c} and \eqref{4.1d} are given by:
\begin{align}
\theta_4\equiv -\frac{\partial \beta_{\lambda^{\rm B}}}{\partial \lambda^{\rm B}} =2-\eta_N^{\rm B}, &\quad\textrm{ and  }\quad
\theta_5\equiv -\frac{\partial\beta_{g^{\rm B}} }{\partial g^{\rm B}} =-2(1+\eta_N^{\rm B}).
\end{align}
At a non-Gaussian fixed point the background anomalous dimension function is $\eta_N^{\rm B}=-2$, establishing that the critical exponents \eqref{thetaback} are 
indeed universal.
\begin{table}[t]
\begin{center}
\begin{tabular}{|c|cc|cc|cc|}
\hline 
Fixed Point & $g_*$  & $\lambda_*$ & $g_*^{\rm B}$ & $\lambda_*^{\rm B}$ & $g_* \lambda_*$ & $g_*^{\rm B} \lambda_*^{\rm B}$ \\ \hline
\small{G-G-FP}      & $0$               & $0$               & $0$   	  & $0$           & $0$             & $0$ \\
\small{G-NG-FP }    & $0$               & $0$               & $2.20$ 	  & $- 0.131$	  & $0$ 	    & $ -0.29 $ \\
\small{NG-G-FP }    & $1.055$            &  $0.222$           &  $0$   	  &  $0$ 	  & $0.234$		& $0$ \\ 
\small{NG-NG-FP }   & $1.055$            &  $0.222$           &  $-41.649$ &  $ 0.578$      & $0.234$		& $-24.06$ \\ \hline
 \end{tabular}
\caption{\small{This table shows all fixed points occurring in the RG flow of the four-parameter system without mixed term  given by \eqref{4.1}. The products $g_* \lambda_*$
and $g_*^{\rm B} \lambda_*^{\rm B}$ are also given.}}
\label{tab:mk=0}
\end{center}
\end{table}

At this stage, it is useful to pause and have a closer look at the general mechanism that generates the NGFP for the background couplings. For a typical background coupling $\bar{u}_k^{\rm B} \int d^4x \sqrt{\gb} \bar{\cO}$ with mass-dimension $d_m$ the $\beta$-function for the corresponding dimensionless coupling $ u_k^{\rm B} = k^{- d_m} \bar{u}_k^{\rm B} $ will assume the form
\be\label{betagen}
\p_t u_k^{\rm B} = - d_m \, u^{\rm B}_k + f(u^{\rm A}_k, u^{\rm mix}_k) \, ,
\ee
where $u^{\rm A}_k$ and $ u^{\rm mix}_k$ denote the dimensionless coupling constants multiplying the interaction terms including the ``genuine'' metric and a mixture of $g$ and $\gb$, respectively. The later are determined through their corresponding $\beta$-functions so that $f(u^{\rm A}_*, u^{\rm mix}_*)$ is a fixed number. The fixed point value for the background coupling is then obtained by solving the linear equation $\beta_{u^{\rm B}_*} = 0$:
\be\label{backfix}
u^{\rm B}_* = (d_m)^{-1} \, f(u^{\rm A}_*, u^{\rm mix}_*) \, .
\ee
Notably, this mechanism works for all {\it dimensionful} background coupling constants, but fails if $u^{\rm B}_k $ is power-counting marginal, i.e. it has $d_m = 0$. In the latter case $u^{\rm B}_k $ does not obtain a finite fixed point value and runs logarithmically in the UV. Clearly, it would be highly desirable to get a better understanding of this very perculiar feature. A complete clarification
of this issue is, however, beyond the scope of the present work and will be left for future study.  

A further question arising naturally at this point is how the bimetric results relate to those coming from the single-metric truncation. 
We must then compare the subsystem \eqref{4.1a}-\eqref{4.1b} to the system of $\beta$-functions arising in a single-metric truncation as given in
\eqref{sm-betafct}. We find that the system \eqref{sm-betafct} has a NGFP located at
\begin{equation}
g_*^{\rm SM} =  \,1.129 , \qquad \lambda_*^{\rm SM} =  0.216 \, .
\end{equation}
Diagonalizing the stability matrix yields to complex conjugate critical exponents with
\be
 \theta_{\rm SM}'=\,1.709,\qquad\theta_{\rm SM}''=\, 3.44 \, .
\ee
These figures refer to exactly the same computational setting (cutoff type, shape function, field parametrization, etc.) as the bimetric computation
above. According to the discussion that led to the identification \eqref{ident} we expect that, if the single metric truncation is a reliable approximation
to the bimetric truncation, we should find $g_*^{\rm SM}\approx g_*$ and $\lambda_*^{\rm SM}\approx \lambda_*$.
Comparing the single metric NGFP with the NG-G-FP
or the NG-NG-FP of Table \ref{tab:mk=0} we see that these relations are satisfied remarkably well.
 However, the critical exponents of the two systems are  rather different. This confirms that the bimetric corrections are indeed important at the
quantitative level.

\begin{figure}[t!]
  \centering
  \subfloat[][\small{Bimetric truncation}]{\label{fig:plot-A}\includegraphics[width=0.4\textwidth]{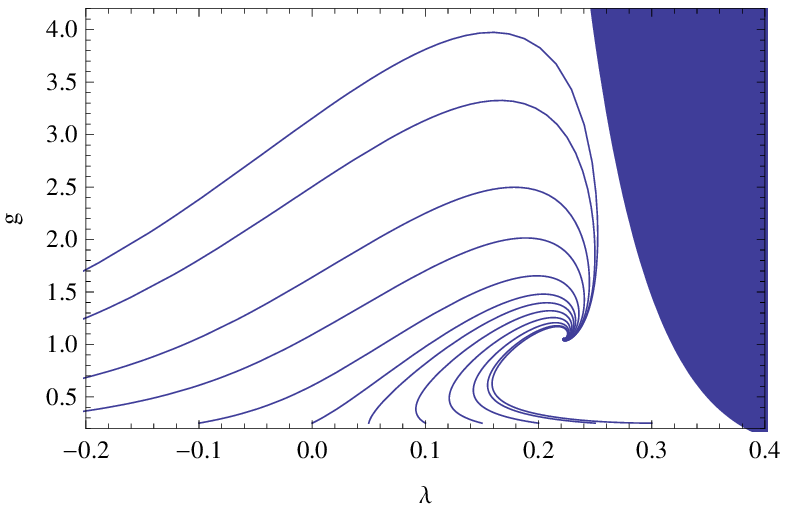}}\hspace{1cm}
  \subfloat[][\small{Single-metric truncation}]{\label{fig:plot-SM}\includegraphics[width=0.4\textwidth]{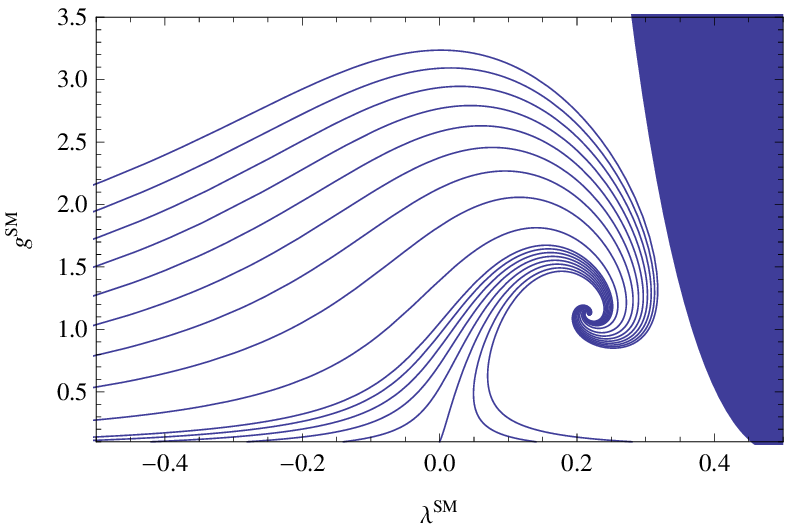}}
 \caption[small]{\small{Phase portrait of the bimetric (left) and single metric truncation (right). The phase portraits show that both systems are
qualitatively very similar. The plots also show the boundary of the physical parameter space.}}
\label{fig:A-SM}
  \end{figure}

We close this subsection by comparing the phase portraits resulting from solving the bimetric \eqref{4.1} and single-metric flow equations \eqref{sm-betafct}. An illustrative set of numerically obtained sample trajectories is shown in Figure  \ref{fig:A-SM}.
Remarkably, both flows exhibit the same qualitative behavior, despite being based on two quite different systems of differential equations.  
In both cases we can distinguish trajectories that run for decreasing $k$ towards negative cosmological constants, referred to as trajectories of Type Ia in the
terminology of \cite{frank1}, to positive cosmological constant (Type IIIa), and to a vanishing $\lambda$. The latter, single trajectory (of Type IIa) is
a separatrix; it crosses-over from the NGFP to the GFP. Both in the single metric and the bimetric case all trajectories in the upper half plane (positive Newton 
constant) are pulled into the non-Gaussian fixed point as we send $k\to\infty$.
The shaded region in the plots of Figure \ref{fig:A-SM} is delimitated by a line where  the anomalous
dimension $\eta_N$ diverges \cite{frank1} and  does not belong to the physical parameter space. In the bimetric case, too, all trajectories of Type IIIa terminate at a finite value of $k$ at this boundary. 
Exactly as in the single-metric case they cannot be continued to the physical point $k=0$ within the truncation used \cite{frank2}. 

%

\subsection{The double-Einstein-Hilbert truncation including mixed interactions}
\label{sect:4.2}

Now we  switch on the mixed interaction term $\propto m_k$ and  search for zeros of  the complete system of $\beta$-functions, that is, for
fixed points
 $\mathbf{g}_*\equiv (g_*, \lambda_*, m_*, g^{\rm B}_*,\lambda^{\rm B}_*)$.
We analyze  \eqref{betaphys4} and \eqref{betaback4} for  general values of the parameter  $n=2,3,4\dots$. Notably, this system has the same 
lower triangular structure as in the case described in Subsection \ref{sect:4.1}. Thus the 
critical exponents corresponding to the background couplings are exactly the same.
However, allowing for  a non-zero value of $m_k$ entails that the NGFP found before now splits into  two different NGFPs, referred to as UV-NGFP and IR-NGFP, respectively.

We now exemplify these structures for the special case $n=2$. The corresponding position of the two fixed points
is given in Table \ref{tab:UVIR}. 
\begin{table}
\begin{center}
\begin{tabular}{r|c|ccc|cc|cc|}
\cline{2-9}
& \multicolumn{1}{|c|} {Fixed Point} & $g_*$  & $\lambda_*$ & $m_*$   & $g_*^{\rm B}$ & $\lambda_*^{\rm B}$ & $g_* \lambda_*$ & $g_*^{\rm B} \lambda_*^{\rm B}$ 
\\ \cline{1-9}
\multicolumn{1}{|c|}{\multirow{2}{*}{$n=2$}} &
\multicolumn{1}{|c|}  {UV-NGFP }    & $ 1.273$ &  $0.237$     & $0.025$ &  $ 2.009 $    &  $ 0.133$ & $0.3017$ & $0.267$   \\ \cline{2-9}
\multicolumn{1}{|c|}{}                        &
\multicolumn{1}{|c|}  {IR-NGFP}    & $0.821$ &  $ 0.210$     & $ - 0.080$ &  $ -1.339$ &  $0.251$     & $0.172$& $-0.336$    \\ \cline{1-9}
\end{tabular}
\caption{\small{Position of the fixed points occurring in the double-Einstein-Hilbert truncation \eqref{gr:ansatz} for  the sample value $n=2$. The products
$g_* \lambda_*$ and $g_*^{\rm B} \lambda_*^{\rm B}$ are also given. }}
\label{tab:UVIR}
\end{center}
\end{table}
Linearizing the $\beta$-functions near the fixed points,
 we find that the
 flow near the UV-NGFP (IR-NGFP) is governed by a pair of complex conjugate critical exponents
$\theta_1=\theta_2^*\equiv \theta'+i\theta''$, a positive (negative) real critical exponent $\theta_3$, and the same universal background critical exponents as before
\begin{subequations}
\begin{align}
\textrm{UV-NGFP:} & &\theta' = 4.800 & & \theta''= 8.722  ,& & \theta_3 = 8.151 , && \theta_4 = 4 , && \theta_5 = 2  \label{mixed_a},\\
\textrm{IR-NGFP:} & &\theta' = 3.131 & & \theta''= 2.821  ,& & \theta_3 = -24.814  , && \theta_4 = 4 ,&&\theta_5 = 2 \label{mixed_b} .
\end{align}
\end{subequations}
These stability properties then motivate the denominations ``UV-NGFP''  for the fixed point whose eigendirections
are all UV-attractive and ``IR-NGFP'' for the one with one UV-repulsive eigendirection, respectively. In fact,
we shall shortly see, that there exists a cross-over trajectory which emanates from the UV-NGFP in the UV and ends at the IR-NGFP in the
limit $k \rightarrow 0$.

\begin{figure}[t] 
  \centering
  \subfloat[][\tiny{UV- and IR-$g_*(n)$ }]{\label{fig:g_n}\includegraphics[width=0.3\textwidth]{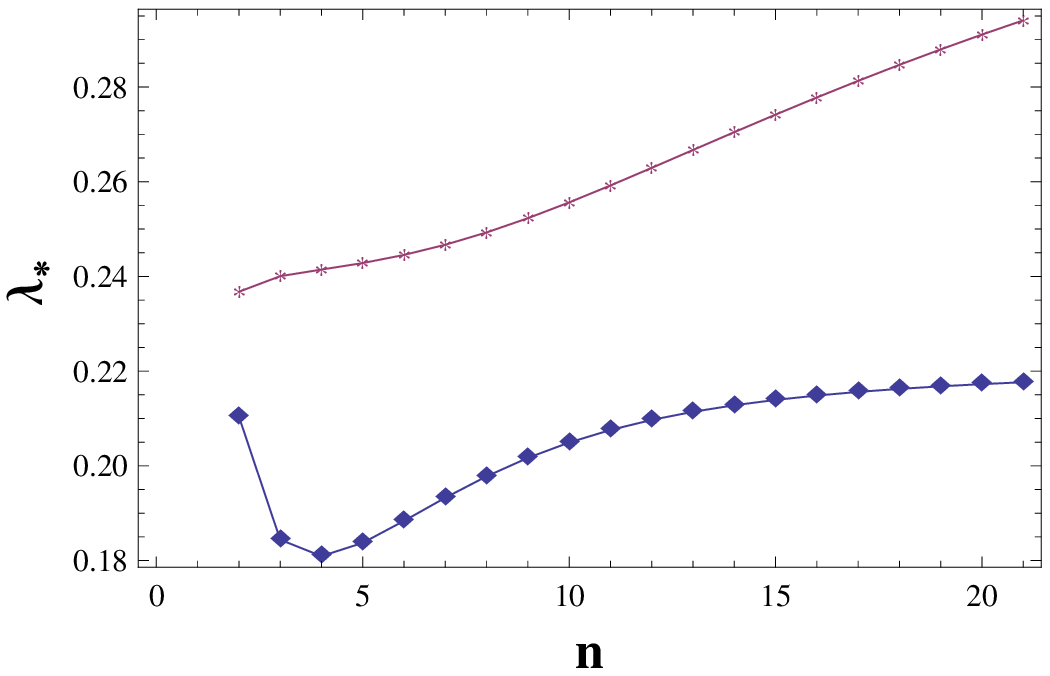}}\hspace{0.3cm}                
  \subfloat[][\tiny{UV- and IR-$\lambda_*(n)$ }]{\label{fig:lambda_n}\includegraphics[width=0.3\textwidth]{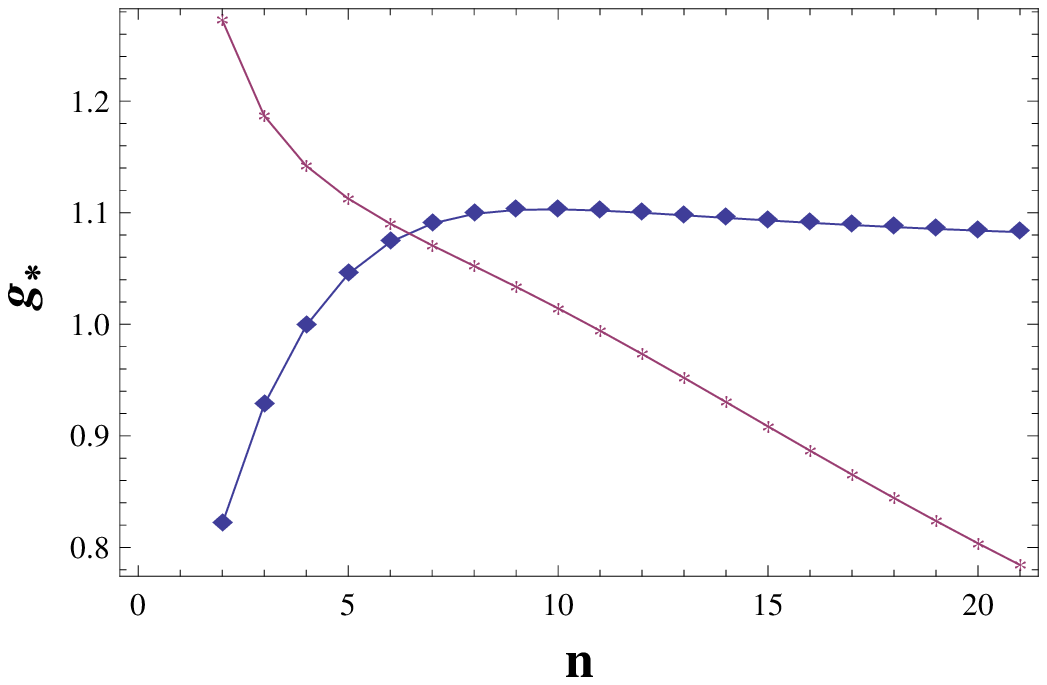}}\hspace{0.3cm} 
  \subfloat[][\tiny{UV- and IR-$m_*(n)$}]{\label{fig:m_n}\includegraphics[width=0.3\textwidth]{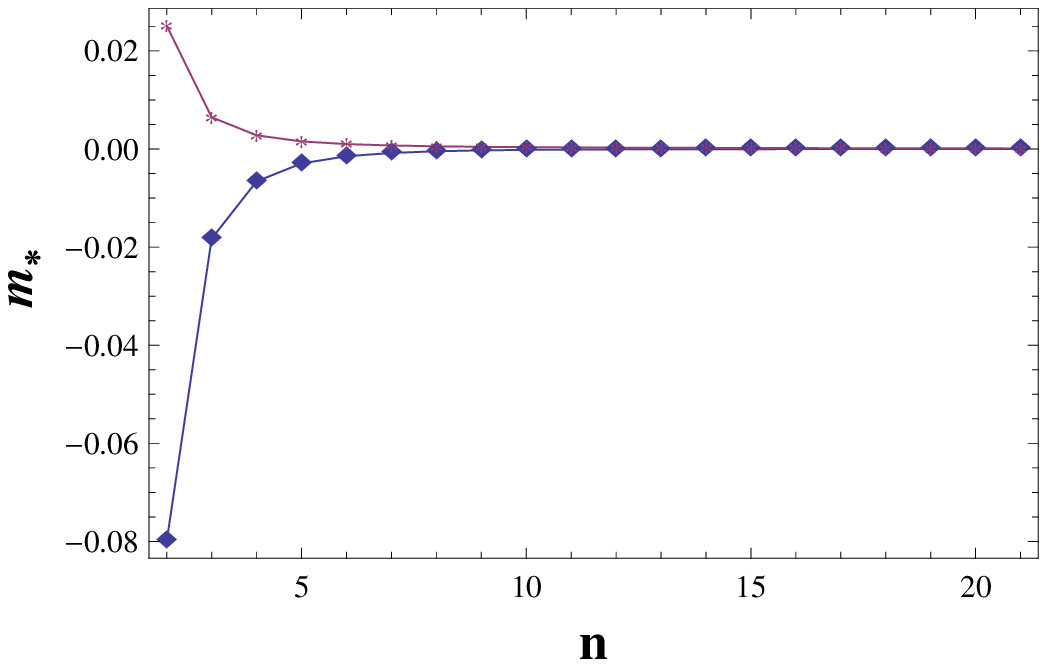}}

  \subfloat[][\tiny{UV- and IR-$\theta'(n)$ }]{\label{fig:thetr*}\includegraphics[width=0.3\textwidth]{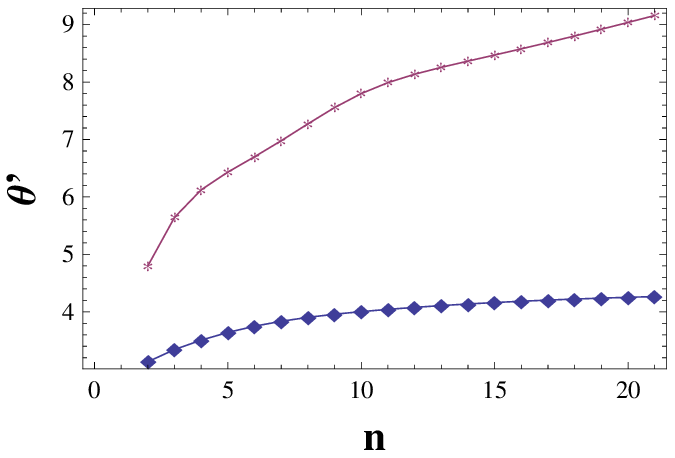}}\hspace{0.3cm}               
  \subfloat[][\tiny{UV- and IR-$\theta''(n)$ }]{\label{fig:theti*}\includegraphics[width=0.3\textwidth]{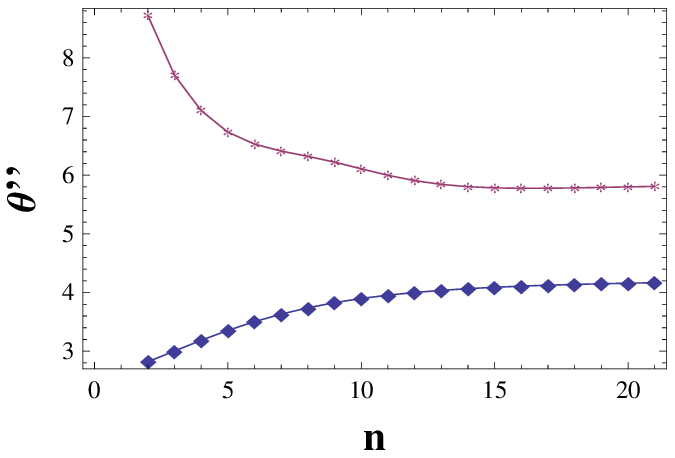}}\hspace{0.3cm}
  \subfloat[][\tiny{UV- and IR-$\theta_3(n)$}]{\label{fig:thet3*}\includegraphics[width=0.3\textwidth]{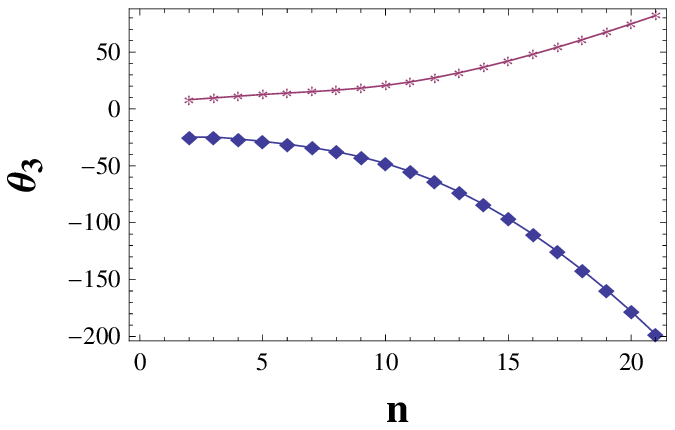}}
\caption{\small{The $n$-dependence of the two NGFP in the $\{g,\lambda,m\}$-subsystem  (a)-(c), together with their
$n$-dependent  stability coefficients (d)-(f).
 The plotted circles represent the IR-NGFP whereas the ``$*$'' represent the UV-NGFP.}}
\label{fig:UV-IR-fp}
 \end{figure}

\begin{figure}[t]
\begin{center}
\subfloat[][\small{Three-dimensional phase portrait  }]{\label{fig:phaseportrait3d}\includegraphics[scale=1.2]{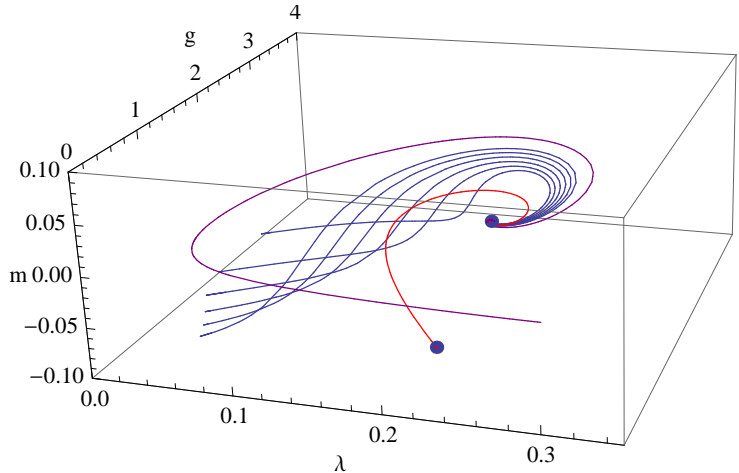} }\hspace{0.6cm}                
\subfloat[][\small{Two-dimensional projection}]{\label{fig:phaseportrait2d}\includegraphics[scale=1.1]{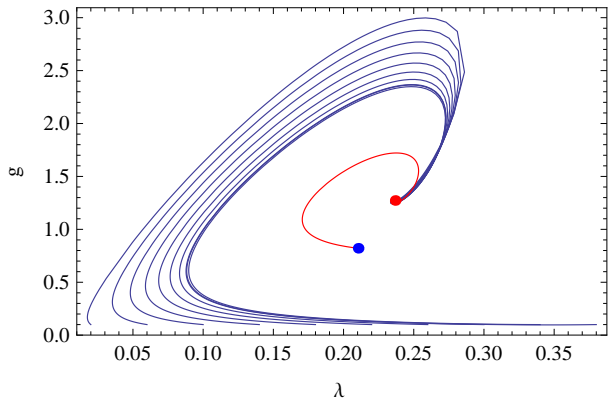} }\hspace{0.6cm} 
\caption[]{\small{(a) Phase portrait of the $\{g,\lambda,m\}$-subsystem.
(b) The projection onto the  $g$-$\lambda-$plane of some sample trajectories. The special trajectory crossing over from the UV-NGFP to the IR-NGFP is shown in both plots.}}
\end{center}
\label{fig:3dplot}
\end{figure}

The position and critical exponents of the UV-NGFP and IR-NGFP for general values $n$ are shown in Figure \ref{fig:UV-IR-fp}. 
Notably, both the UV-NGFP and IR-NGFP occur for all values $n \ge 2$. 
Interestingly, the fixed
point value $m_*$ tends rapidly to zero as we increase the value of  $n$, in both cases (cf.\ Figure \ref{fig:m_n}).
In Figure \ref{fig:phaseportrait3d} we show the 3-dimensional phase diagram of the  $\{g, \lambda, m\}$-subsystem with its UV-NGFP and IR-NGFP.
The sample trajectories in this plot have been obtained numerically.
Figure \ref{fig:phaseportrait2d}  depicts the projection of the 3-dimensional flow onto the $g$-$\lambda-$plane. Remarkably, this projection 
looks very similar to the flow generated by the single metric Einstein-Hilbert truncation shown in Figure \ref{fig:plot-SM}.  
The Figure \ref{fig:phaseportrait3d} allows us to identify the 3-dimensional generalizations of the familiar trajectories of Type Ia and IIIa, i.e. their ``lift'' to
$g$-$\lambda$-$m-$space. The $g$-$\lambda-$projection of the 3-dimensional trajectories displayed is almost identical to what one obtains from the single metric 
truncation. In particular, according to both the 3-dimensional and 2-dimensional system of RG equations the trajectories of Type IIIa terminate at finite $k$ in a boundary singularity.
For both the 3-dimensional and the 2-dimensional single-metric RG equations this class contains trajectories whose turning point (the point where $\beta_{\lambda}=0$)
is arbitrarily close to the GFP.  Those trajectories spend a very long RG time near the GFP. As a result, their termination at the boundary singularity can be 
deferred until very late RG times, that is, small scale $k$. All these features common to the 2- and 3-dimensional RG flow fortify the rather impressive robustness of the projected phase portrait.
In refs. \cite{h3} and \cite{entropy} it has been pointed out that if one uses a trajectory of Type IIIa to define the quantum theory, the mechanism of ``using up''  a
long RG time near the GFP is instrumental in obtaining a theory with a classical regime given by General Relativity where $G_k = {\rm constant}$ for a long interval of scales $k$.
The same argument also applies to the ``lifted'' trajectories on $g$-$\lambda$-$m-$space. For a tentative matching of these model trajectories to the
observations made in  Nature the arguments in \cite{h3} and \cite{entropy} basically remain unaltered therefore.

A highly intriguing new feature of the 3-dimensional flow is the emergence of a novel IR-NGFP and, as a result, the cross-over trajectory connecting the UV-NGFP to the IR-NGFP. It emanates from 
the UV-NGFP in the UV and for $k\to 0$ approaches the IR-NGFP along its only IR attractive direction, see Figures \ref{fig:phaseportrait3d} and \ref{fig:phaseportrait2d}.
For no value of $k$ this trajectory gets close to the GFP. As a result, it does {\it not} give rise to a classical regime: the dimensionful parameters
$G_k^{\rm A},\, \Lambda_k^{\rm A},\,M_k$ have a significant scale dependence all the way down from ``$k=\infty$'' to $k=0$.

While the cross-over trajectory in the form obtained within the double-Einstein-Hilbert truncation cannot be used as a model for the real world, its existence and, more generally, 
the emergence of an IR fixed point, is a highly welcome feature of the new truncation. One reason is that, for the first time, we have found a trajectory
which has positive $\lambda$ everywhere {\it and which does not terminate in a singularity}. As to yet it had always been unclear what kind of average action could
avoid the singularity near $\lambda=1/2$. (See \cite{frank2} for an early attempt at solving this problem.) Moreover an IR fixed point, albeit at a 
different location, had been argued to lead to a particularly realistic late-time cosmology and an explanation of the
``recent'' cosmic acceleration \cite{cosmo2}.

Clearly, for phenomenological purposes  the IR-NGFP is not yet satisfactory. But the fact that it occurs at all in the new truncation is very encouraging. It shows that we are ``on the right track'' towards a better understanding of the infrared.

\section{Discussion and Conclusions}
\label{sect:5}

In this paper we analyzed the functional RG equation for full-fledged Quantum Einstein Gravity within the framework of a novel class of approximations,
the bimetric truncations of theory space. In the ansatz for the effective average action we included two different Einstein-Hilbert actions, one for the dynamical
and another for the background metric, as well as a non-derivative term which mixes the two. Our main interest was in assessing whether the resulting RG flow lends itself
to the Asymptotic Safety construction of a microscopic theory of quantized gravity. One key result is that as far as one can tell within the restrictions of the
approximation, the answer is clearly positive. In particular, we discovered a bimetric generalization of the non-Gaussian fixed point
known to exist within the single-metric Einstein-Hilbert truncation. This confirms earlier expectations \cite{elisa2} originating from a similar bimetric
computation within conformally reduced gravity. Moreover, we also found a second non-trivial fixed point which might control the theory's IR behavior.

The main technical innovation of the present paper is the conformal projection technique. As its main virtue
it allows to probe (a part of) the bimetric gravitational theory space without the necessity of evaluating complicated operator
traces involving the two Laplacians constructed from both the ``genuine'' metric $g$ and the background metric $\gb$, $\Tr[f(-\Db^2, -D^2)]$. 
A complementary approach could be based on expanding $\Gamma_k[h; \gb]$ in terms of the $n$-point functions of $h_{\m\n}\equiv g_{\m\n}-\gb_{\m\n}$. 
Results for full-fledged quantum gravity using the latter approach will be reported elsewhere \cite{alcod}.

It is impressive to compare the $\beta$-functions for the couplings $g, \lambda$, \eqref{4.1a} and  \eqref{4.1b} to the corresponding ones arising in the single
metric computation, eqs.\ \eqref{sm-betafct}. The latter use the same gauge fixing and auxiliary field construction, so that this can serve as an illustration 
which makes the new features
of the bimetric computation  transparent. Obviously, the $\beta$-functions \eqref{4.1} and the single metric result \eqref{sm-betafct} are {\it very} different 
in their analytical structure.
In particular {\it all coefficients} multiplying threshold functions, as well as their pole structure at $\l=1/2$ and $\l=3/4$, are manifestly different.
 In this light it is miraculous that the single-metric and bimetric treatments, nevertheless,  give rise to a non-Gaussian fixed point {\it with very similar properties}.
For this reason we may be optimistic that the bimetric generalizations of the RG flows we know already will, at least qualitatively, confirm the essential
features of the single-metric truncations. This concerns in particular the viability of the Asymptotic Safety program.

\section*{Acknowledgments}
%
We thank A.\ Codello and J.\ Pawlowski for interesting discussions. The research of E.M.\ and F.S.\ is supported by the Deutsche Forschungsgemeinschaft (DFG)
within the Emmy-Noether program (Grant SA/1975 1-1).

\begin{appendix}
\section{Threshold functions}
\label{App:A}
In this appendix, we collect various definitions and review the central properties of the threshold functions used in the main text. 
In order to evaluate the trace contributions, we use the early time expansion of the heat kernel. For the operators appearing in Table \ref{t.1} this can be done utilizing the the ``master formula'' \cite{mr}
\begin{equation}\label{Tr:exp}
\Tr_s\left[W(-\Db^2)\right] = \tfrac{1}{(4 \pi)^{d/2}} \tr_s(\unit_s) \left\{ Q_{d/2}[W] \int d^dx \sqrt{\gb} + \frac{1}{6} Q_{d/2-1}[W] \int d^dx \sqrt{\gb} \Rb + \cdots \right\} \, , 
\end{equation}
where
\begin{equation}\label{Qndef}
Q_n[W] \equiv \frac{1}{\Gamma(n)} \int_0^\infty \rmd z z^{n-1} W(z) \, ,
\end{equation}
and the higher-order terms outside the truncation are indicated by the dots.
The subscript $s = {\rm 2T}, {\rm 1T}, 0$ indicates that the Laplacians act on transverse-traceless symmetric tensors, transverse vectors, and scalars, respectively. 
The algebraic trace  $\tr_s(\unit_s)$ counts the degrees of freedom in the corresponding sector. Setting $d_s = \tr_s(\unit_s)$ we have
\be\label{def:mult}
d_{\rm 2T} = \tfrac{1}{2} (d-2)(d+1)  \, , \qquad d_{\rm 1T} = d-1  \, , \qquad d_{0} = 1 \, .
\ee
When performing the double-expansion of the traces \eqref{flowev} in $\eps, \Rb$, the Q-functionals appearing in the trace-evaluation can be related to
the standard threshold functions via
\be\label{Qev}
Q_n\left[ \frac{z^q \, \p_t (Z_k R_k)}{Z_k \, (z + R_k + wk^2)^p} \right] \,
= \tfrac{\Gamma(n+q)}{\Gamma(n)} k^{2(n+q-p+1)} \, \left[ 2 \, \Phi^p_{n+q}(w) - \eta_N \, \Pt^p_{n+q}(w) \right] \, , \; n+q > 0 \, .
\ee
Here $\eta_N = - \p_t \ln Z_k$ and $z = -\Db^2$. The cutoff-scheme dependence is encoded in the standard dimensionless threshold functions \cite{mr}
\be\label{phidef}
\begin{split}
\Phi^p_n(w) \equiv & \, \frac{1}{\Gamma(n)} \int^\infty_0 dz z^{n-1} \frac{R^{(0)}(z) - z R^{(0)\prime}(z)}{\Big[z + R^{(0)}(z) + w\Big]^p} \, , \\
\Pt^p_n(w) \equiv & \, \frac{1}{\Gamma(n)} \int^\infty_0 dz z^{n-1} \frac{R^{(0)}(z)}{\Big[z + R^{(0)}(z) + w\Big]^p} \, . 
\end{split}
\ee
To simplify the notation of the traces including the running Newton's constant $Z_k$, the right-hand-side of \eqref{Qev} suggests
introducing
\be\label{qdef}
q_n^p(w)\=\Phi^p_n(w)-\tfrac{1}{2}\eta_N\;\widetilde{\Phi}^p_n(w) \, .
\ee
As we shall see later on, this short-hand notation will considerably simplify the notational complexity of the evaluated traces.
 
For the explicit evaluation of the flow equations, we resort to the optimized  cutoff \cite{opt}. In this case
the shape-function entering into \eqref{phidef} is given by 
\be
R^{(0)}(z) = (1-z)\theta\,(1-z) \, 
\ee
and the integrals appearing in the threshold functions can be carried out analytically:
\be
\Phi^p_n(w) = \frac{1}{\Gamma(n+1)} \, \frac{1}{(1+w)^p} \, , \quad \Pt^p_n(w) = \frac{1}{\Gamma(n+2)} \, \frac{1}{(1+w)^p} \, .
\ee
We exclusively resort to this type of cutoff in the main part of the paper.

\section{Double-Einstein-Hilbert truncation: deriving the $\beta$-functions}
\label{App:B}
In this appendix we present the technical details entering into the evaluation of the truncated flow equation \eqref{flowev}. We start with the explicit calculation of the operator traces constructed from the entries of Table \ref{t.1} using the intermediate formulas reviewed in Appendix \ref{App:A}. Subsequently, the projection of the RG flow gives rise to our main result, the $d$-dimensional $\beta$-functions of the double-Einstein Hilbert truncation, \eqref{betaphys} and \eqref{betaback} below.
 
Our starting point are the operator traces entering into \eqref{flowev}. Here we first focus on the gravitational sector sourced by the transverse-traceless $h_{\m\n}^{\rm T}$ and  the trace $h$, before evaluating the universal contributions from the gauge- and auxiliary sectors given by the second and third block of Table \ref{t.1}, respectively. All expressions are given in terms of the dimensionless coupling constants
\be\label{dimless}
\l_k = k^{-2} \Lambda_k^{\rm A} \, , \quad g_k = k^{d-2} G_k^{\rm A} \, , \quad m_k = k^{-2} M_k \, , \quad \lambda^{\rm B}_k = k^{-2} \Lambda^{\rm B}_k \, , \quad g^{\rm B}_k = k^{d-2} G^{\rm B}_k \, .
\ee

Comparing the first and second line of Table \ref{t.1}, we observe that the expressions entering into $\cS_0$ and $\cS_{\rm 2T}$ are structurally very similar. This motivates the following helicity-dependent definitions, in terms of which the evaluated traces take the same structural form. First, setting $\ell = 1$, we identify the arguments that will enter into the dimensionless threshold functions as the $\Delta$- and $\Rb$-independent parts of the square brackets in Table \ref{t.1},
\be\label{arg}
 w_{\rm 2T} =  \tilde{c}_{\rm 2T} \, m_k - 2 \l_k  \, , \qquad w_0 =  \tilde{c}_0 \, m_k - \tfrac{d}{d-1} \l_k \, . 
\ee
The $(d,n)$-dependent constants $\tilde{c}_{\rm 2T}$ and $ \tilde{c}_0$ given in \eqref{csconst}. Furthermore, the $\epsilon$-expansion motivates introducing short-hand expressions for the $\epsilon$-derivatives of the $m, \lambda$-dependent terms in Table \ref{t.1}
\be\label{argder}
\begin{array}{ll}
w_{\rm 2T}^\prime = \tilde{c}_{\rm 2T} \, \alpha_2 \, m_k - 2 \, \alpha_0 \, \lambda_k  \, , 
&w_0^\prime =  \tilde{c}_0 \, \alpha_2 \, m_k - \tfrac{d}{d-1} \alpha_0 \lambda_k    \, , \\[1.1ex]
w_{\rm 2T}^{\prime\prime} =  \tilde{c}_{\rm 2T} \alpha_2 (\alpha_2-1) m_k - 2 \alpha_0(\alpha_0-1) \lambda_k  \, , 
& w_{\rm 0}^{\prime\prime} =  \tilde{c}_{\rm 0} \alpha_2 (\alpha_2-1) m_k - \tfrac{d}{d-1} \alpha_0(\alpha_0-1) \lambda_k \, ,
\end{array}
\ee
with $\alpha_i$ defined in \eqref{alphaeq}.
In terms of these $\cS_0$ and $\cS_{\rm 2T}$ assume the same, though spin-dependent, form. The explicit computation, utilizing the formulas in Appendix \ref{App:A} yields
\be\label{Sgrev}
\begin{split}
\cS_s = & \, \tfrac{d_s}{(4 \pi)^{d/2}} \, q^1_{d/2}(w_s) \,  k^d  \int d^dx \sqrt{\gb} \\ & \,
 + \tfrac{d_s}{(4 \pi)^{d/2}} \, \left[ \tfrac{1}{6} q^1_{d/2-1}(w_s) - C_s q^2_{d/2}(w_s) \right] k^{d-2} \int d^dx \sqrt{\gb} \Rb \\ & \, 
 - \tfrac{d_s}{(4 \pi)^{d/2}} \, \left[ \tfrac{\nu d(d-6)}{4} q^2_{d/2+1}(w_s) +  w_s^\prime \, q^2_{d/2}(w_s) \right] k^{d} \int d^dx \sqrt{\gb} \eps \\ & \,
 - \tfrac{d_s}{(4 \pi)^{d/2}} \Big[  \tfrac{\nu(d-6)}{2} ( \tfrac{d-2}{12}+ C_s )  q^2_{d/2}(w_s) + \tfrac{1}{6} \, w_s^\prime \, q^2_{d/2-1}(w_s) \\
& \qquad \qquad  \quad - \tfrac{C_s}{2} \left( \nu d(d-6) q^3_{d/2+1}(w_s) + 4 \, w_s^\prime \, q^3_{d/2}(w_s) \right) \Big] k^{d-2} \int d^dx \sqrt{\gb} \Rb \eps \\ & \,
+ \tfrac{d_s}{(4 \pi)^{d/2}} \, \Big[
\tfrac{\nu^2 d (d+2)}{16}  (d-6)^2 q^3_{d/2+2}(w_s)
+ \tfrac{\nu d }{2}  (d-6) w_s^\prime q^3_{d/2+1}(w_s) +  (w_s^\prime)^2 q^3_{d/2}(w_s) \\ & \qquad \qquad \quad 
- \tfrac{\nu (d-6)d}{16}((d-6)\nu-2) \, q^2_{d/2+1}(w_s) 
- \tfrac{1}{2} w_s^{\prime\prime} \, q^2_{d/2}(w_s)
\Big]  k^{d} \int d^dx \sqrt{\gb} \eps^2 \, .
\end{split}
\ee
The gauge-sector, given by the second block in Table \ref{t.1}, likewise leads to
\begin{align}\label{Sgfev}
\cS_{\rm gf} = & \, - \tfrac{1}{(4\pi)^{d/2}} \Big[ \Phi^1_{d/2} k^d \int d^dx \sqrt{\gb} + \tfrac{1}{6} \Phi^1_{d/2-1} k^{d-2} \int d^dx \sqrt{\gb} \Rb \big] \\ & \,
+ \tfrac{\nu d}{(4\pi)^{d/2}} \Big[ (d+2) \Phi^3_{d/2+2} + (d-1) \Phi^2_{d/2+1} \Big] k^d  \int d^dx \sqrt{\gb} \eps\nonumber \\
& \, + \tfrac{\nu }{(4\pi)^{d/2}} 
\Big[ \tfrac{(d-1)(d^2-2d-12)}{6d} \Phi^2_{d/2} + \tfrac{d^3+9d^2-34d+12}{6(d-1)} \Phi^3_{d/2+1} + \tfrac{3d(d+2)}{2(d-1)} \Phi^4_{d/2+2}
\Big] k^{d-2} \int d^dx \sqrt{\gb} \Rb \eps\nonumber \\
& - \tfrac{d \nu}{2(4\pi)^{d/2}}  \Big[ (1-\nu) (d-1) \Phi^2_{d/2+1} - (1-2\nu+d\nu)(d+2) \Phi^3_{d/2+2} \nonumber\\
& \qquad \qquad + \half \nu (d+6)(d+4)(d+2) \Phi^5_{d/2+4}\Big] k^d  \int d^dx \sqrt{\gb} \eps^2 \,\nonumber .
\end{align}
Here all the threshold functions are evaluated at $w = 0$, and we dropped the argument for notational simplicity.

The auxiliary fields contribute to the running of the background couplings only. Their trace is universal, i.e. it does not depend on any coupling constant. It reads
\be\label{Sauxev}
\begin{split}
\cS_{\rm aux} =   - \tfrac{1}{(4\pi)^{d/2}} k^d \int d^dx \sqrt{\gb} \Big[ & \, 
(d-1)  \Phi^1_{d/2}  
+ \left( \tfrac{d-1}{6} \Phi^1_{d/2-1} + \tfrac{d^2-d+1}{d(d-1)} \Phi^2_{d/2} \right)  \Rb k^{-2}
\Big] \, . 
\end{split}
\ee
Also here, all threshold functions are evaluated at zero argument, which is dropped for simplicity.

When constructing the $\beta$-functions from the ansatz \eqref{EAAA}, it turns out to be useful 
to organize the right-hand-side of the flow equation in terms of the following
double expansion in $\Rb$, $\eps$:
\be\label{Fexp}
\left. \p_t \Gamma_k[g, \gb] \right|_{g = (1+\epsilon)^{\nu} \gb} =\sum_{k,l} \, F^{(k,l)}(g_k, \lambda_k, m_k) \, k^d \int d^dx \sqrt{\gb} \,  \epsilon^k (\Rb k^{-2})^l \, .
\ee
The dimensionless expansion coefficients $F^{(k,l)}$ depend parametrically on $n, d$ as well as on the couplings $g_k, \lambda_k$ and $m_k$. They are,
however,  independent 
of the background coupling constants $g_k^{\rm B}, \lambda_k^{\rm B}$. Their explicit form is easily read off by substituting \eqref{Sgrev}, \eqref{Sgfev}, and \eqref{Sauxev} into
\eqref{flowev} and comparing to the expansion \eqref{flowev}. We find 
\be
\begin{split}\label{f00-f10}
F^{(0,0)}= &  \frac{1}{(4\pi)^{d/2}}\left[ d_{\rm 2T} \, q^1_{d/2}(w_{\rm 2T}) + q^1_{d/2}(w_0) - d \Phi_{d/2}^1 \right] \, , \\
F^{(0,1)}= & \frac{1}{(4\pi)^{d/2}} \Big[d_{\rm 2T}\,\Big\{\tfrac{1}{6}q^1_{d/2-1}(w_{\rm 2T})-C_{\rm T}q^2_{d/2}(w_{\rm 2T})\Big\}+
 \tfrac{1}{6}q^1_{d/2-1}(w_{0})-C_{S}q^2_{d/2}(w_0)\\
{}& \qquad \qquad -\tfrac{d^2-d+1}{d(d-1)} \Phi_{d/2}^2-\tfrac{d}{6}\Phi_{d/2-1}^1
 \Big] \, , \\
 F^{(1,0)} = & \frac{1}{(4\pi)^{d/2}} \Big[
  - d_{\rm 2T} \, w_{\rm 2T}^{\prime} \, q^2_{d/2}(w_{\rm 2T}) - w_0^{\prime} \, q^2_{d/2}(w_0) \\
 &  
 - \tfrac{\nu d (d-6)}{4}  \left\{ d_{\rm 2T} q^2_{d/2+1}(w_{\rm 2T}) + q^2_{d/2+1}(w_0)\right\} 
 + \nu d(d+2) \Phi^3_{d/2+2} + \nu d(d-1) \Phi^2_{d/2+1} \Big] \, ,
\end{split}
\ee
together with
\be
\begin{split}\label{f11}
F^{(1,1)} = & \frac{1}{(4\pi)^{d/2}} \Big[ 
- \tfrac{\nu(d-6)}{2} \left\{ d_{\rm 2T} ( \tfrac{d-2}{12} + C_{T} ) q^2_{d/2}(w_{\rm 2T}) + ( \tfrac{d-2}{12} + C_{S} ) q^2_{d/2}(w_{\rm 0}) \right\} \\
& - \tfrac{1}{6} \left\{ d_{\rm 2T} \, w_{\rm 2T}^{\prime} \, q^2_{d/2-1}(w_{\rm 2T}) +  w_{\rm 0}^{\prime} \, q^2_{d/2-1}(w_{\rm 0}) \right\} 
+2 d_{\rm 2T} C_T \, w_{\rm 2T}^{\prime} q^3_{d/2}(w_{\rm 2T})  \\
& +2 C_S w_0^{\prime} q^3_{d/2}(w_0) + \tfrac{\nu d (d-6)}{2} \left\{ d_{\rm 2T} C_T q^3_{d/2+1}(w_{\rm 2T}) + C_S q^3_{d/2+1}(w_{\rm 0}) \right\} 
\\
& + \tfrac{\nu(d-1)(d^2-2d-12)}{6d} \Phi^2_{d/2} + \nu \tfrac{d^3+9d^2-34d+12}{6(d-1)} \Phi^3_{d/2+1} + \nu \tfrac{3d(d+2)}{2(d-1)} \Phi^4_{d/2+2}
\Big] \, ,
\end{split}
\ee
\be
\begin{split}\label{f20}
F^{(2,0)} = & \frac{1}{(4\pi)^{d/2}} \Big[ 
\tfrac{\nu^2 d(d+2)}{16} (d-6)^2 \left\{ d_{\rm 2T} q^3_{d/2+2}(w_{\rm 2T}) + q^3_{d/2+2}(w_{\rm 0}) \right\}
\\ &
+ \tfrac{ \nu d(d-6)}{2}  \left\{ d_{\rm 2T} \, w_{\rm 2T}^{\prime} \,  q^3_{d/2+1}(w_{\rm 2T}) + w_{\rm 0}^{\prime} \, q^3_{d/2+1}(w_{\rm 0}) \right\} 
\\ &
+  d_{\rm 2T} \, (w_{\rm 2T}^{\prime})^2 \,  q^3_{d/2}(w_{\rm 2T}) + (w_{\rm 0}^{\prime})^2 \, q^3_{d/2}(w_{\rm 0})  
- \half \left\{ d_{\rm 2T} w_{\rm 2T}^{\prime\prime} \, q_{d/2}^2(w_{\rm 2T}) + w_{\rm 0}^{\prime\prime} \, q_{d/2}^2(w_{\rm 0}) \right\}
\\ &
- \tfrac{\nu d(d-6)}{16} ((d-6) \nu -2) \left\{  d_{\rm 2T} q^2_{d/2+1}(w_{\rm 2T}) + q^2_{d/2+1}(w_{\rm 0}) \right\}
\\ &
- \tfrac{\nu d(1-\nu)(d-1)}{2}  \Phi^2_{d/2+1} - \tfrac{d\nu (1-2\nu + d \nu) (d+2)}{2} \Phi^3_{d/2+2} - \tfrac{d\nu^2 (d+6)(d+4)(d+2)}{4}  \Phi^5_{d/2+4}
\Big] \, . 
\end{split}
\ee
Note the implicit $\eta_N$-dependence contained in $q^p_n(w)$.

Substituting \eqref{LHSflow} into \eqref{Fexp} we  obtain a system of coupled differential equations governing the
RG dependence of the dimensionless couplings \eqref{dimless}. The anomalous dimension for Newton's constant is found solving the linear equation
\be\label{etafct}
\eta_N = 16 \pi g_k F^{(1,1)} \, 
\ee
with $ F^{(1,1)}$ given by \eqref{f11}, for $\eta_N$. The result has the structure
\be
\eta_N=\frac{g_k\;B_1(\lambda_k,m_k)}{1-g_k\;B_2(\lambda_k,\m_k)} \, ,
\ee
where the functions $B_1$ and $B_2$ can be read off from \eqref{f11} by splitting the expression for $F^{(1,1)}$ into the terms independent of, and linear in
$\eta_N$ (as in \cite{mr}).
 
Based on this solution of $\eta_N$, the $\beta$-functions for the dimensionless $g_k, \lambda_k$ and $m_k$ close among themselves:
\be\label{betaphys}
\begin{split}
\beta_g = & \left[\, d-2 + \eta_N \, \right]g_k \, , \\
\beta_{\l} = & (\eta_N - 2) \lambda_k + 8 \pi g_k \, \tfrac{(d-2)}{ n d^2} \,  \left[ 2 (d-2) F^{(2,0)} - (2-nd) F^{(1,0)} \right] \, , \\
\beta_{m} = & (\eta_N - 2) m_k + 16 \pi g_k \, \tfrac{(d-2)}{ n (1-n)d^2} \,  \left[ (d-2) F^{(2,0)} - F^{(1,0)} \right] \, .
\end{split}
\ee
The $\beta$-functions governing the running of the background couplings are
\be\label{betaback}
\begin{split}
\beta_{g^{\rm B}} =&  \left[ d-2 + \eta_N^{\rm B} \right]  \, g^{\rm B}_k \, , \\
\beta_{\lambda^{\rm B}} = & (\eta_N^{\rm B} - 2) \lambda_k^{\rm B} + 8 \pi g_k^{\rm B} \, \tfrac{d-2}{(1-n)d^2} \left[ \tfrac{(1-n)d^2}{d-2} F^{(0,0)} + 2 (d-2) F^{(2,0)} - (d+2-nd) F^{(1,0)} \right] \, .
\end{split}
\ee
The background anomalous dimension id linear in $g_k^{\rm B}$ and given by
\be\label{etafctB}
\eta_N^{\rm B} = 16 \pi \left( F^{(0,1)} - F^{(1,1)} \right) g_k^{\rm B} \, . 
\ee

This completes the derivation of the $\beta$-functions of the bimetric Einstein-Hilbert truncation including the non-trivial mixed term proportional to $m_k$. They are valid for arbitrary space-time dimension $d$, and constitute the central result of this appendix. Restricting to $d=4$, their properties are discussed in Section \ref{sect:4}. 

We close this appendix by recovering the $\beta$-functions for the double-Einstein-Hilbert truncation without mixed terms $\propto M_k$, eq.\ \eqref{4.1}. These readily follow from the general result \eqref{betaphys} and \eqref{betaback} by setting $m_k = 0$ and eliminating $F^{(2,0)}$ via the constraint
\be\label{B.15}
\left. (d-2) F^{(2,0)} - F^{(1,0)} \right|_{m_k = 0} = 0 \, ,
\ee
which results from the consistency condition $\beta_m|_{m=0} = 0$. Restricting to $d=4$, one then arrives at the $\beta$-functions \eqref{4.1} analyzed in Subsection \ref{sect:4.1}.

\end{appendix}

\end{document}